\documentclass[journal=jpcafh]{achemso}
\setkeys{acs}{articletitle = true}
\SectionNumbersOn

\usepackage[version=3]{mhchem} % Formula subscripts using \ce{}
\usepackage{amsmath,amssymb,amsfonts}
\usepackage{xcolor}
\usepackage{longtable}
\usepackage{bm}
\usepackage{array}
\usepackage{lscape}
\usepackage[colorlinks]{hyperref}
\hypersetup{ % mute the intense default colors for links
    colorlinks=true,
    linkcolor=red!80!black,
    urlcolor=magenta!80!black,
    citecolor=green!70!black,
}
\usepackage[nameinlink, capitalize]{cleveref}
\usepackage[utf8]{inputenc}
\usepackage[T1]{fontenc}
\usepackage{braket}
\usepackage{caption}
\usepackage{subcaption}
\usepackage{graphicx}
\usepackage{wrapfig}
\usepackage{booktabs}

\newcommand{\ee}[1]{\mathrm{e}^{#1}}

\title{Chemical Interpretation of Time-Dependent Coupled-Cluster Theory}
\author{Aparna Krishnan}
\affiliation[VT]{Department of Chemistry, Virginia Tech, Blacksburg, VA 24061, USA}
\author{H{\aa}kon Emil Kristiansen}
\affiliation[UiO]{Hylleraas Centre for Quantum Molecular Sciences, Department of Chemistry, University of Oslo, P.O. Box 1033 Blindern, N-0315 Oslo, Norway}
\author{Benjamin G. Peyton}
\affiliation[VT]{Department of Chemistry, Michigan State University, East Lansing, MI 48824, USA}
\author{T. Daniel Crawford}
\affiliation[VT]{Department of Chemistry, Virginia Tech, Blacksburg, VA 24061, USA}
\email{crawdad@vt.edu}
\author{Thomas Bondo Pedersen}
\affiliation[UiO]{Hylleraas Centre for Quantum Molecular Sciences, Department of Chemistry, University of Oslo, P.O. Box 1033 Blindern, N-0315 Oslo, Norway}
\email{t.b.pedersen@kjemi.uio.no}

\begin{document}

\section*{Abstract}

While providing a highly accurate framework for simulating laser-induced many-electron dynamics in atom and molecules, including linear and nonlinear
steady-state and transient absorption spectra, time-dependent coupled-cluster theory does not offer a straightforward interpretation in chemical terms.
This should be contrasted with conventional time-independent equation-of-motion coupled-cluster or frequency-dependent response models where a simple
eigenvector analysis readily reveals the dominant orbital-excitation character of individual excited states.
We fill this gap by expanding the left and right coupled-cluster functions in Slater-determinant basis, thus allowing for a time-dependent
generalization of configuration weights that can be used to track populations throughout a simulation. The same expansions are
used to decompose the time-dependent electric-dipole moment and autocorrelation function, providing a computationally
straightforward approach to the assignment of absorption peaks to orbital transitions for single-reference systems.
At the time-dependent coupled-cluster singles-and-doubles level of theory, we demonstrate the power of the proposed methodology
by assigning valence lines in the linear absorption spectra of four ten-electron molecules (\ce{HF}, \ce{H2O}, \ce{NH3}, and \ce{CH4})
with different point-group symmetries, validating the assignment by comparison with equation-of-motion coupled-cluster singles-and-doubles theory. In addition, core-level excitations are assigned for \ce{HF}, \ce{H2O}, and \ce{NH3}.
Finally, the usefulness of time-dependent configuration weights is illustrated by applications to an impulsive stimulated X-ray
Raman scattering process in the \ce{Ne} atom and to a transient pump-probe spectrum of the \ce{HF} molecule.

\section{Introduction} \label{sec:intro}

The accurate identification and analysis of electronic excitations are fundamental to understanding how molecules respond to external electromagnetic fields, particularly in spectroscopic applications\cite{suzuki2012electronic,helgaker2013molecular}. While traditional approaches focus on equilibrium properties in the energy domain, real-time methods have emerged as a powerful tool for simulating the full time evolution of a system under various external perturbations such as lasers\cite{english2015perspectives,goings2018real,palacios_quantum_2020,li2020real}. This dynamic approach enables the study of non-equilibrium processes\cite{kolesov2016real} and ultrafast phenomena, including electron dynamics and photo-induced reactions. The real-time propagation of the system's wavefunction provides a direct route to simulating transient processes and calculating time-dependent molecular properties.

Among these methods, real-time time-dependent coupled-cluster (TD-CC) theory\cite{Hoodbhoy1979,Arponen1983,Pedersen1998,huber2011explicitly,Kvaal2012,nascimento2016linear,Sato2018,pedersen2019symplectic,Hansen2019,Koulias2019,skeidsvoll2020time,wang2022accelerating,sverdrup2023time,kvaal_time-dependent_2025} is a robust framework for simulating electronic transitions by tracking the time evolution of a correlated wavefunction. The primary way to obtain spectral information is by calculating time-dependent expectation values, for example multipole moments such as the electric dipole moment, and performing a Fourier transformation to yield the desired spectrum. In practice, the discrete Fourier transform is used, requiring a very large number of time steps (say, $100\,000$ steps or more) to achieve sufficient spectral resolution. Since each time step typically demands a computational effort comparable to a few ground-state CC iterations, it is essential to not only develop accelerated algorithms for each time step but also to minimize the number of time steps required through accurate extrapolation of induced quantities. 
While graphical processing units show some potential for significant acceleration of TD-CC simulations\cite{wang2022accelerating,wang_real-time_2025},
new algorithms that exploit sparsity and locality of electron correlation are needed\cite{Crawford2019,peyton2023reduced}.
Dipole extrapolation techniques have recently been developed to minimize the number of time steps required to achieve high spectral resolution
or, for weak field kicks, even infinite resolution\cite{hauge2021extrapolating,hauge_cost-efficient_2023,kick_super-resolution_2024}.

Although such acceleration techniques must be further developed, another challenge has barely been addressed to date: How to extract chemical insight from TD-CC simulations? Spectral lines may of course be assigned to molecular-orbital (MO) contributions by identifying the corresponding equation-of-motion coupled-cluster
(EOM-CC)\cite{stanton1993equation}
eigenvector through the transition frequency, or the population of EOM-CC eigenstates may be tracked during the dynamics as proposed by \citet{Pedersen2021}
The obvious downside of both approaches is that the time-independent EOM-CC problem must be solved in addition to running the TD-CC simulations. 

Alternatives do exist to extract information directly from a simulation, however. The autocorrelation function (ACF) provides information about
the energy and weight of the excited states that participate in the dynamics,\cite{pedersen2019symplectic}
and configuration weights---the population of individual Slater determinants in the CC state, introduced in Ref.~\citenum{configurationwts}
as a bivariational extension of the cluster expansion analysis by \citet{cizek_cluster_1969} for CC ground states---may be computed at each point in time, providing a picture of orbital transitions in real time. Finally, one may decompose expectation values into
individual MO contributions whose Fourier transform reveals their spectral contributions.

Decomposition methods have been used earlier to analyze and interpret complex spectra \cite{wibowo2021modeling}. \citet{repisky2015excitation} introduced a dipole-weighted transition matrix analysis that facilitates the interpretation of spectral features by viewing the absorption spectrum as a sum of contributions from ground-state MO pairs.
This decomposition provides valuable insight into the occupied-virtual orbital transitions involved in each spectral feature.
Building on this work, \citet{bruner2016accelerated} used a similar transition dipole decomposition not only for interpretation but also for accelerating spectral calculations. 
By first decomposing the total dipole into individual MO pair contributions, they were able to use Pad\'{e} approximants to converge the spectrum from significantly shorter simulation times, making RT methods more competitive with traditional linear-response approaches, especially for large systems with a high density of states.
These methods, however, predominantly rely on time-dependent density functional theory (TD-DFT) and a transition dipole-based approach.
Our work extends a closely related idea by performing a dual analysis that compares the insights from both dipole and ACF decompositions.
Preliminary work\cite{hauge2021extrapolating} indicate that a straightforward decomposition of the TD-CC one-body reduced density matrix 
does not consistently provide identification of orbital transitions in agreement with EOM-CC theory.
We ascribe this observation to the ``convolution'' of elementary processes built into the density matrix and propose an alternative
formulation that aligns well with the linear excitation operator formalism of EOM-CC theory.

More recently, \citet{Abraham2026} developed a component analysis of the time-dependent double coupled-cluster (TD-dCC) Green's function that decomposes core-level ionization spectra into direct and hole-mediated excitation pathways, 
demonstrating that orbital-resolved decomposition within TD-CC frameworks provides direct access to many-body satellite formation mechanisms. While their approach targets ionization spectra via propagation of a correlated $(N-1)$-electron state, the present work pursues a complementary decomposition strategy for neutral absorption spectra through direct analysis of the time-evolved ground-state amplitudes.

The dipole decomposition partitions the time-dependent dipole moment into contributions from individual orbital excitation channels while ACF decomposition works by projecting the autocorrelation function onto individual orbital excitations, thus providing a frequency-resolved, orbital-resolved picture of the system's dynamic response. The ACF decomposition is more sensitive to nonlinear effects, as it captures how electronic states are populated and how their contributions evolve over time. Furthermore, through this framework, we can obtain information about forbidden transitions and can also investigate doubly excited states in contrast to dipole decomposition where only single-excitation contributions are accessible.  Through both the dipole and ACF frameworks, we can obtain information about orbital contributions to individual absorptions, but the latter also provides information about forbidden transitions and can also investigate doubly excited states.

To validate the physical insights gained from these decompositions, we systematically compare our findings with stationary-state results from EOM-CC singles and doubles (EOM-CCSD) theory.
By confirming that the dominant peaks in our decomposed spectra correspond to EOM-CCSD transitions predictions---in both energy and orbital character---we establish a crucial link between the dynamic and static descriptions of molecular excitation. This integrated approach allows us to investigate how static and dynamic descriptions can complement each other, offering a comprehensive and interpretable framework for analyzing time-dependent spectroscopy using correlated electronic structure methods.

Three fundamental questions motivate this work. First, do dipole and ACF decomposition methods, based on ground-state transition moments versus time-evolved populations, identify the same orbital characters for electronic excitations? Second, given that molecular symmetry governs directional field selectivity of absorptions, can we gain additional insight into the MO contributions to electronic transitions using symmetry-resolved probes? Third, when do these methods provide complementary rather than redundant information? We address these questions through systematic TD-CCSD calculations on four ten-electron species: \ce{HF} ($C_{\infty v}$), \ce{H2O} ($C_{2v}$), \ce{NH3} ($C_{3v}$), and \ce{CH4} ($T_d$). For the first three molecules, we also assign core-level excitations, extending the framework to the X-ray regime.
Furthermore, we investigate how time-dependent configuration weights can be used to gain insights into the MO mechanism of impulsive stimulated X-ray Raman scattering (ISXRS) in the \ce{Ne} atom and, in combination with dipole decomposition, into a transient pump-probe spectrum of the \ce{HF} molecule.

In this work, we limit the scope to TD-CC models where the underlying orbital space is independent of time.
Such ``conventional'' TD-CC models are known to become unstable in situations where the laser fields drive the dynamics so far from equilibrium
that the weight of the reference determinant approaches zero \cite{pedersen2019symplectic}. To describe such dynamics, the orbitals must be allowed to evolve in time in concert with the correlating cluster amplitudes \cite{Kvaal2012,Sato2018,kristiansen2020numerical}.
The formalism developed in the present paper can in principle be straightforwardly applied to such orbital-adaptive methods as well.
In practice, however, the computational effort required to compute overlaps between determinants in different (not biorthonormal) orbital
bases necessitates the development of approximation schemes. This is beyond the scope of the present work and is deferred to future work.

The remainder of this paper is organized as follows: In Section \ref{sec:theory}, we present the theoretical framework for time-dependent coupled cluster theory and give expressions for configuration weights, dipole, and ACF decompositions. Section \ref{sec:results} contains our results for linear valence absorption
spectra of the \ce{HF}, \ce{H2O}, \ce{NH3}, and \ce{CH4} molecules, the ISXRS process in the \ce{Ne} atom, and a transient pump-probe spectrum of the \ce{HF} molecule. 
Finally, Section \ref{sec:conc} summarizes our key findings and their implications for practical analysis of TD-CC simulations.

\section{Theory} \label{sec:theory}

\subsection{Time-dependent coupled cluster theory}
The non-relativistic time-dependent Schr\"{o}dinger equation (TDSE) governs the evolution of a quantum system over time\cite{scrinzi2014time} and can be written as (in atomic units (a.u.))
\begin{equation}\label{tdse}
    \mathrm{i} \frac{\partial}{\partial t} \Psi(t) = \hat{H}(t) \Psi(t)
\end{equation}
where \( t \) denotes time, \( \Psi(t) \) is the time-dependent wave function of the system, \( \hat{H}(t) \) is the time-dependent Hamiltonian operator where in this work the time-dependence arises from an oscillating external electric field,
\begin{equation}
    \hat{H}(t) = \hat{H}_0 - \hat{\boldsymbol{d}} \cdot \boldsymbol{E}(t)
\end{equation}
Here, we take $\hat{H}_0$ to be the clamped-nuclei electronic Hamiltonian,
$\hat{\boldsymbol{d}}$ is the electric-dipole operator, and $\boldsymbol{E}(t)$ is the external electric field.
The bivariational framework of CC theory introduces independent approximations of the wave function and its conjugate\cite{Arponen1983,sverdrup2023time,kvaal_time-dependent_2025}. The CC bra and ket are defined as\cite{helgaker2013molecular}
\begin{equation}
    \langle\tilde{\Psi}(t)| = \langle{\Phi_0}|\hat{\Lambda}(t)\mathrm{e}^{-\hat{T}(t)}
\end{equation}
and 
\begin{equation}
    |\Psi (t)\rangle = \mathrm{e}^{\hat{T}(t)} |\Phi_0\rangle
\end{equation}
and they together represent the quantum state of a many-electron system\cite{pedersen2019symplectic}.
The cluster operators $\hat{T}$ and $\hat{\Lambda}$ are defined as
\begin{equation}
    \hat{T}(t) = \sum_{\mu = 0}^{N} t_\mu (t) \hat{X}_\mu
\end{equation}
and 
\begin{equation}
    \hat{\Lambda}(t) = \sum_{\mu = 0}^{N} \hat{X}_\mu^{\dagger} \lambda_\mu (t)
\end{equation}
where $\hat{X}_\mu$ is an excitation operator that maps the reference determinant $\Phi_0$ to the excited determinant $\Phi_\mu$,
\begin{equation}
    |\Phi_\mu\rangle = \hat{X}_\mu |\Phi_0\rangle
\end{equation}
with $\hat{X}_0 = \hat{I}$ (the identity operator).
These determinants constitute an orthonormal system
\begin{equation}
    \langle\Phi_\mu|\Phi_\nu\rangle = \braket{\Phi_0 \vert \hat{X}_\mu^\dagger X_\nu \vert \Phi_0} = \delta_{\mu\nu}, \qquad \mu,\nu \geq 0
\end{equation}
The time-dependent bivariational principle
leads to equations of motion for the cluster amplitudes $t_\mu (t)$ and $\lambda_\mu (t)$\cite{pedersen2019symplectic}:
\begin{align}
    &\mathrm{i}\frac{\partial{t_{\mu}(t)}}{\partial{t}} = \langle\Phi_{\mu}|\mathrm{e}^{-\hat{T}(t)}\hat{H}(t)|{\Psi(t)}\rangle \\
    -&\mathrm{i}\frac{\partial{\lambda_{\mu}(t)}}{\partial{t}} = \langle\tilde{\Psi}(t)|[\hat{H}(t), \hat{X}_{\mu}]|\Psi(t)\rangle
\end{align}
Note that $\lambda_0$ is constant and will be chosen equal to unity such that $\braket{\tilde{\Psi}(t) \vert \Psi(t)} = 1$ for all $t$.
These equations reduce to the ground-state CC equations in the time-independent limit, with $t_0(t) = -\mathrm{i}\mathcal{E}_0 t$
where $\mathcal{E}_0$ is the CC ground-state energy.

In most practical applications, the TDSE is solved by numerically propagating the wave function forward in time from a given initial state, usually the ground state in the case of electron dynamics. This approach forms the foundation of real-time methods, where the wave function's time evolution is tracked under various perturbations. In the TD-CC case, time-dependent observables may then be computed as expectation values\cite{pedersen2019symplectic},
\begin{equation}
    \Omega(t) = \mathrm{Re}\braket{\tilde{\Psi}(t) \vert \hat{\Omega} \vert \Psi(t)}
\end{equation}
where $\hat{\Omega}$ is a Hermitian operator. Fourier transformation of TD-CC expectation values provides information about interference
between the (unkown) stationary states of the system but does not directly yield chemical insight related to molecular orbitals.
The following sections discuss how that information may be extracted from TD-CC expectation values and from the evolving TD-CC quantum state itself.

\subsection{Configuration weights in TD-CC theory}

In Ref.~\citenum{configurationwts}, \citeauthor{configurationwts} introduced configuration weights as expectation values of projection
operators onto individual Slater determinants, providing a simple analysis of time-independent CC states in configuration-interaction-like
terms. In the present work, we generalize the definition of configuration weights to TD-CC states:
\begin{equation}
    W_{\mu}(t) = \mathrm{Re}\,\tilde{c}_\mu(t) c_\mu(t)
\end{equation}
where $\mu \geq 0$, and
\begin{equation}
    \tilde{c}_\mu(t) = \braket{\tilde{\Psi}(t) \vert \Phi_\mu}, \qquad
    c_\mu(t) = \braket{\Phi_\mu \vert \Psi(t)}
\end{equation}
Explicit expressions (in spin-orbital basis) for these coefficients up to triples can be found in Ref.~\citenum{configurationwts}.

The reference weight is given by
\begin{equation}
    W_0(t) =  \mathrm{Re}\,\tilde{c}_0(t) c_0(t)
\end{equation}
while the total singles, doubles, etc. weights are obtained by summation over subsets of the excited Slater determinants,
\begin{equation}
    W_1(t) = \sum_\mu^\mathrm{singles} W_\mu(t), \qquad
    W_2(t) = \sum_\mu^\mathrm{doubles} W_\mu(t), \qquad \mathrm{etc.}
\end{equation}
The weights are normalized in the sense
\begin{equation}
    \sum_{\mu\geq 0} W_\mu(t) = \sum_{I=0}^{N_e} W_I(t) = 1
\end{equation}
where $N_e$ is the number of electrons.
However, due to the linear parameterization of the de-excitation cluster operator $\hat{\Lambda}$,
the configuration weights vanish above the truncation level of the cluster operators. For the 
CC singles and doubles (CCSD) model, for example, $W_I = 0$ for $I > 2$---i.e., CC models behave in
much the same way as truncated configuration-interaction states\cite{configurationwts}.
The important exception is the lack of strict bounds. The bivariational formulation implies that
CC weights are not mathematically guaranteed to be non-negative and at most unity\cite{configurationwts}.

Complementary to stationary-state populations~\cite{Pedersen2021},
time-dependent configuration weights can directly provide a chemically appealing interpretation of
laser-driven many-electron processes such as impulsive stimulated X-ray Raman scattering in terms
of orbital excitations and de-excitations. Importantly, this approach does not rely on explicit
calculation of excited states using EOM-CC theory but, of course, the weights must change sufficiently during the
matter-field interaction to sustain such a real-time analysis. For correlated dynamics driven by weak fields,
a different approach is required, such as dipole decomposition.

\subsection{Time-dependent dipole moment and its orbital decomposition}

A key application of TD-CC methods is the calculation of absorption spectra.
The absorption cross section $\sigma(\omega)$ can be expressed in terms of the spectral response function $S(\omega)$ 
and the total field energy per unit area $I(\omega)$ at frequency $\omega$ as
\begin{equation}
    \sigma(\omega) = \frac{\omega S(\omega)}{I(\omega)}
\end{equation}
The spectral response function, in turn, can be obtained from a spectral analysis of the work done by the field
on the electronic system~\cite{aurbakken_transient_2024}. Within the electric-dipole approximation, the spectral
response function reads~\cite{aurbakken_transient_2024}
\begin{equation}
    \label{eq:spectral_response_function}
    S(\omega) = -2\mathrm{Im} \left[\tilde{\boldsymbol{d}}(\omega) \cdot \tilde{\boldsymbol{E}}(-\omega)\right]
\end{equation}
where $\tilde{\boldsymbol{d}}(\omega)$ and $\tilde{\boldsymbol{E}}(\omega)$ are the Fourier transforms\cite{bracewell1989fourier}
of the electric-dipole moment $\boldsymbol{d}(t) = \mathrm{Re}\braket{\tilde{\Psi}(t) \vert \hat{\boldsymbol{d}} \vert \Psi(t)}$
and of the electric field, respectively,
\begin{align}
    &\tilde{\boldsymbol{d}}(\omega) = \frac{1}{\sqrt{2\pi}} \int_{-\infty}^{\infty} \mathrm{d}t\, \boldsymbol{d}(t) \ee{\mathrm{i}\omega t} \\
    &\tilde{\boldsymbol{E}}(\omega) = \frac{1}{\sqrt{2\pi}} \int_{-\infty}^{\infty} \mathrm{d}t\, \boldsymbol{E}(t) \ee{\mathrm{i}\omega t}
\end{align}
Note that the spectral response function and the cross section are a signed quantities, positive when the electronic system absorbs energy and negative when energy is released.

This semi-classical formulation of absorption spectroscopy is applicable regardless of the shape of the external electric field and thus
can be used to compute both linear and nonlinear absorption spectra, including transient pump-probe spectra.
To compute a spectrum, the system is subjected to a short external perturbation (electric field), inducing a time-dependent dipole response, 
$\boldsymbol{d}(t)$, which is recorded as the system evolves. This is then Fourier transformed\cite{bracewell1989fourier} and the resulting frequency-domain dipole moment, \( \tilde{\boldsymbol{d}}(\omega) \), reveals the system's absorption properties. The peaks in the absorption spectrum correspond to the excitation energies of the system, while the intensities of these peaks are related to the square of transition dipole moments, providing information on the strength of the transitions. This method allows for the direct computation of excitation energies and transition strengths in a fully time-resolved manner.

For a broadband external field, TD-CC treats all excited states simultaneously, including core excitations. Hence, the time-dependent dipole moment obtained from an RT simulation can be decomposed into orbital contributions to gain insight into individual excitation channels. The time-dependent dipole moment may be written as 
\begin{equation}
    \label{eq:d_decomp}
    \boldsymbol{d}(t) = \mathrm{Re} \langle\tilde{\Psi}(t)|\boldsymbol{\hat{d}}|\Psi(t)\rangle
    = \sum_{\mu =0}^{N}\sum_{\nu=0}^{M} \mathrm{Re}\,\tilde{c}_\mu(t)\langle\Phi_{\mu}|\boldsymbol{\hat{d}}|\Phi_{\nu}\rangle c_{\nu}(t)
\end{equation}
using the decomposition of bra and ket states as
\begin{align}
    |\Psi(t)\rangle &= \sum_{\mu=0}^{M} |\Phi_\mu\rangle c_\mu (t)\\
    \langle\tilde{\Psi}(t)| &= \sum_{\mu=0}^{N} \tilde{c}_\mu (t) \langle\Phi_\mu|
\end{align}
where
$M$ is the total number of determinants that can be generated with the chosen spin-orbital basis and $N$ is the number of suitably ordered excitations/de-excitations included in the cluster operators. $N$ is generally smaller than $M$ and in cases where $N$ is equal to $M$, the left- and right-hand wave functions ($\langle\tilde{\Psi}(t)|$ and $|\Psi(t)\rangle$) are equivalent to the FCI wave function and its hermitian conjugate up to a normalization constant. Note that since $\langle\tilde{\Psi}(t)|\Phi_\mu\rangle = 0$ for any $\mu >N$, the expansion of the bra truncates at $N$ while no such truncation is obtained for the expansion of the ket.

Since $\hat{\boldsymbol{d}}$ is a one-electron operator, the summation of terms stops at one excitation level higher than the coupled-cluster truncation. In the case of a CCSD state, this means the summation includes contributions up to the triples level. Organizing the terms by excitation level---where $0$ refers to the reference state, $S$ to singles, $D$ to doubles, and so forth---we can express the dipole moment by summing these contributions from each excitation. For the CCSD case, it may be written as 
\begin{equation}
    \boldsymbol{d}(t) = \boldsymbol{d}^{00}(t) + \boldsymbol{d}^{0S}(t) + \boldsymbol{d}^{SS}(t) + \boldsymbol{d}^{SD}(t) + \boldsymbol{d}^{DD}(t) + \boldsymbol{d}^{DT}(t)
\end{equation}
where superscripts denote excitation levels in the bra $\langle \tilde \Psi (t)|$ and ket $|\Psi (t)\rangle$ wavefunctions as mentioned above.
For example, $\boldsymbol{d}^{0S}$ contains contributions where the bra is projected onto the reference and the ket onto singles (Eq.~\eqref{eq:d_decomp}), while $\boldsymbol{d}^{SS}$ involves singles in both bra and ket. 

As mentioned earlier, the goal is to identify the dominant orbital contributions to each peak in the TD-CC absorption spectrum. Provided that the CC ground state is dominated by the reference determinant, it should be feasible to identify these transitions by focusing solely on the $0S$ contribution,
\begin{align}
    \boldsymbol{d}^{0S}(t) &= \sum_{ia} \mathrm{Re}\,\langle\Phi_0|\hat{\boldsymbol{d}}|\Phi_{i}^{a}\rangle\tilde{c}_0(t)c_{i}^{a}(t) + \sum_{ia} \mathrm{Re}\,\langle\Phi_{i}^{a}|\hat{\boldsymbol{d}}|\Phi_{0}\rangle\tilde c_{a}^{i}(t)c_{0}(t)\nonumber \\
        &= \sum_{ia} \mathrm{Re}\,\langle{\psi_i}|\hat{\boldsymbol{d}}|\psi_a\rangle[\tilde{c}_0(t)c_{i}^{a}(t) + \tilde{c}_{a}^{i}(t)c_0(t)]
        = \sum_{ia}\boldsymbol{d}_{i}^{a}(t)
\end{align}
Here and in the following, we use indices $i,j$ and $a,b$ to denote occupied and virtual (spin) orbitals.
Focusing on the $0S$ contribution is well justified for not too strong laser fields. Assuming that the CC ground state is dominated by the reference determinant ($W_0 \approx 1$), the $0S$ component captures the primary single-excitation character of each transition.
The remaining terms ($00$, $SS$, $SD$, $DD$, $DT$) contribute with significantly smaller intensities, as confirmed by our decomposition analysis below. In practical implementation, we compute each $\boldsymbol{d}_i^a (t)$ component using the expressions for the reference and single-excitation coefficients provided in Ref.~\citenum{configurationwts} with time-evolved cluster amplitudes. Fourier transformation then yields
frequency-resolved contributions from each occupied-virtual pair $i,a$.

As is clear from the underlying theory, the same frequencies may be found in several components, ${\boldsymbol{d}_i^{a}}$, and their corresponding partial spectra may contain negative peaks. Only the full spectrum---i.e, the sum of the components---is expected to contain positive peaks exclusively in the linear-response regime.

\subsection{Time autocorrelation function and its orbital decomposition}

An alternative analysis to dipole decomposition is based on the ACF, which is defined as the overlap of the wave function with itself at different
points in time. Using both bra and ket to represent the quantum state of the system,
\begin{equation}
    \vert S \rangle\!\rangle = \frac{1}{\sqrt{2}}
    \begin{pmatrix}
        \ket{\Psi} \\
        \ket{\tilde{\Psi}}
    \end{pmatrix}
\end{equation}
and using the indefinite inner product definition of \citeauthor{pedersen2019symplectic}\cite{pedersen2019symplectic},
\begin{equation}
    \langle\!\langle S_1 \vert S_2 \rangle\!\rangle =
    \frac{1}{2} \left(
    \braket{\tilde{\Psi}_1 \vert \Psi_2} + \braket{\tilde{\Psi}_2 \vert \Psi_1}^*
    \right)
\end{equation}
we can write the ACF as
\begin{align}
    A(t^\prime,t) &= \langle\!\langle S(t^\prime) \vert S(t) \rangle\!\rangle \nonumber \\
    &= \frac{1}{2} \braket{\tilde{\Psi}(t^\prime) \vert \Psi(t)}
     + \frac{1}{2} \braket{\tilde{\Psi}(t) \vert \Psi(t^\prime)}^*
\end{align}

If $t^\prime$ is chosen to be the time at which the laser pulse is switched off, the ACF for $t>t^\prime$ contains information
about the stationary states participating in the dynamics, both the total energy of the states and their population.
The ACF was first used for this purpose within TD-CC theory in Ref.~\citenum{pedersen2019symplectic}, to which we refer for further details.
For TD-CCSD, we can expand $|\Psi(t)\rangle$ and $\langle\tilde\Psi(t)|$ using the definitions given above and derive the singles and doubles contribution to the ACF. 
Excluding time-dependent phase factors, the spin-adapted singles and doubles components of the autocorrelation function for a closed-shell reference may be expressed as
\begin{equation}
    \label{eq:acf_a1}
    A_i^a(t^\prime, t) = \frac{1}{2}\bigg(\bigg(\lambda_a^i(t^\prime) - \sum_{jb}\lambda_{ab}^{ij}(t^\prime)t_j^b(t^\prime)\bigg)t_i^a(t) + \bigg(\lambda_a^i(t) - \sum_{jb}\lambda_{ab}^{ij}(t)t_j^b(t)\bigg)^*t_i^{a}(t^\prime)^*\bigg)
\end{equation}
and
\begin{equation}
    \label{eq:acf_a2}
    A_{ij}^{ab}(t^\prime, t) = \frac{1}{2}\bigg(\lambda_{ab}^{ij}(t^\prime)\big(t_{ij}^{ab}(t) + t_i^a(t)t_j^b(t)\big) + \lambda_{ab}^{ij}(t)^*\big(t_{ij}^{ab}(t^\prime) + t_i^a(t^\prime)t_j^b(t^\prime)\big)^*\bigg)
\end{equation}
The fully closed diagrams of the ACF decomposition can also be written as
\begin{align}
A_0(t^\prime, t) &= \frac{1}{2}\bigg(\bigg(1 - \sum_{ia}\lambda_a^i(t^\prime)t_i^a(t^\prime) - \frac{1}{2}\sum_{ijab}\lambda_{ab}^{ij}(t^\prime)t_{ij}^{ab}(t^\prime)
+ \frac{1}{2}\sum_{ijab}\lambda_{ab}^{ij}(t^\prime)t_i^a(t^\prime)t_j^b(t^\prime)\bigg) \nonumber \\
& \qquad + \bigg(1 - \sum_{ia}\lambda_a^i(t)t_i^a(t) - \frac{1}{2}\sum_{ijab}\lambda_{ab}^{ij}(t)t_{ij}^{ab}(t) + \frac{1}{2}\sum_{ijab}\lambda_{ab}^{ij}(t)t_i^a(t)t_j^b(t)\bigg)^*\bigg)
\label{eq:acf_a0}
\end{align}

Equations \eqref{eq:acf_a1}-\eqref{eq:acf_a0} provide the ACF decomposition at the CCSD level, analogous to the dipole decomposition (Eq.~\eqref{eq:d_decomp}) but independent of any particular property operator.  The key distinction is that ACF components, e.g. $A^a_i(t^\prime,t)$) measure the overlap between the wavefunction at reference time $t^\prime$ and its time-evolved form at time $t>t^\prime$ for each excitation channel, encoding both population and phase information. This population-dependent analysis provides sensitivity to field-driven dynamics particularly important for transitions with small dipole moments but significant population transfer under external perturbations.  The time-dependent decomposition of the ACF may be Fourier transformed to the frequency domain to examine the MO contributions to each peak, including dipole-forbidden transitions. 

To enable quantitative comparison between the ACF decomposition and the EOM-CCSD eigenvectors, we apply the following renormalization scheme.
The Fourier transformed ACF components $\tilde{A}^a_i$ and $\tilde{A}^{ab}_{ij}$ at a given frequency represent unitless time-domain amplitudes, while EOM-CCSD eigenvectors follow the intermediate normalization convention in the singles-doubles space. To place both on comparable footing, we normalize the ACF components such that the sum of squared singles and doubles amplitudes equals unity, matching the EOM-CCSD convention. We can obtain the $\tilde{A}_i^a$ and $\tilde{A}_{ij}^{ab}$ values for a particular transition frequency and renormalize them according to
\begin{equation}
    (\tilde{A}_i^a)_\mathrm{norm}(\omega) = N(\omega) \tilde{A}_i^a(\omega)
\end{equation}
and
\begin{equation}
    (\tilde{A}_{ij}^{ab})_\mathrm{norm}(\omega) = N(\omega)\tilde{A}_{ij}^{ab}(\omega)
\end{equation}
with
\begin{equation}
    \label{eq:acf_n}
    N(\omega) = \frac{1}{\sqrt{2\sum_{ia}{(\tilde{A}_i^a(\omega))}^2 + \sum_{ijab} (2\tilde{A}_{ij}^{ab}(\omega) - \tilde{A}_{ij}^{ba}(\omega))\tilde{A}_{ij}^{ab}(\omega)}}
\end{equation}
This normalization establishes a semi-quantitative bridge between dynamic (ACF) and 
static (EOM-CCSD) descriptions, enabling direct comparison of orbital character. 
We note that perfect numerical agreement is not expected as the ACF measures time-evolved population dynamics while EOM-CCSD provides stationary-state eigenvector components, but the normalized amplitudes should exhibit consistent trends and identify the same dominant orbital contributions.

\section{Results} \label{sec:results}

\subsection{Linear absorption spectra}

To test the performance of the dipole and ACF decompositions for linear absorption spectra, we analyze the response of the ten-electron molecules, \ce{HF}, \ce{H2O}, \ce{NH3} and \ce{CH4} using TD-CCSD and the cc-pVDZ basis set\cite{Dunning1989} to a time-dependent delta pulse
\begin{equation}
    \boldsymbol{E}(t) = \boldsymbol{u}E_0 \delta(t)
\end{equation}
where $\boldsymbol{u}$ is a real unit polarization vector along one of the Cartesian axes, and $E_0 = 0.01\,\mathrm{a.u.}$ is the electric-field strength.
While the cc-pVDZ basis set is too small to yield converged linear absorption spectra, it is sufficient to illustrate the
usability of the dipole and ACF decompositions. An augmented and core-valence correlated basis set is used for transient
absorption spectroscopy below.
We propagate the TD-CCSD state for $1000\,\mathrm{a.u.}$ with a time step $\Delta t = 0.01\,\mathrm{a.u.}$ using the fourth-order Runge-Kutta (RK4) integrator. The time step is chosen to ensure stability and accuracy of the RK4 propagation, and the $1000\,\text{a.u.}$ propagation time provides sufficient spectral resolution ($\sim\!0.0063\,\mathrm{a.u.}$) for the present purposes. 
We perform the simulations with the PyCC\cite{pycc} Python program built on top of the
Psi4\cite{smith2020psi4} open-source quantum chemistry package. In our simulations the field kick is switched on at $t = 0.02\,\mathrm{a.u.}$, and all electrons are correlated in the TD-CCSD simulations.
The molecular geometries are provided in the Supporting Information.

\subsubsection{Hydrogen Fluoride}

\begin{figure}[htbp]
 \centering
 \vspace*{5pt}%
  \begin{subfigure}{0.5\textwidth}
    \centering
    \includegraphics[width=\textwidth]{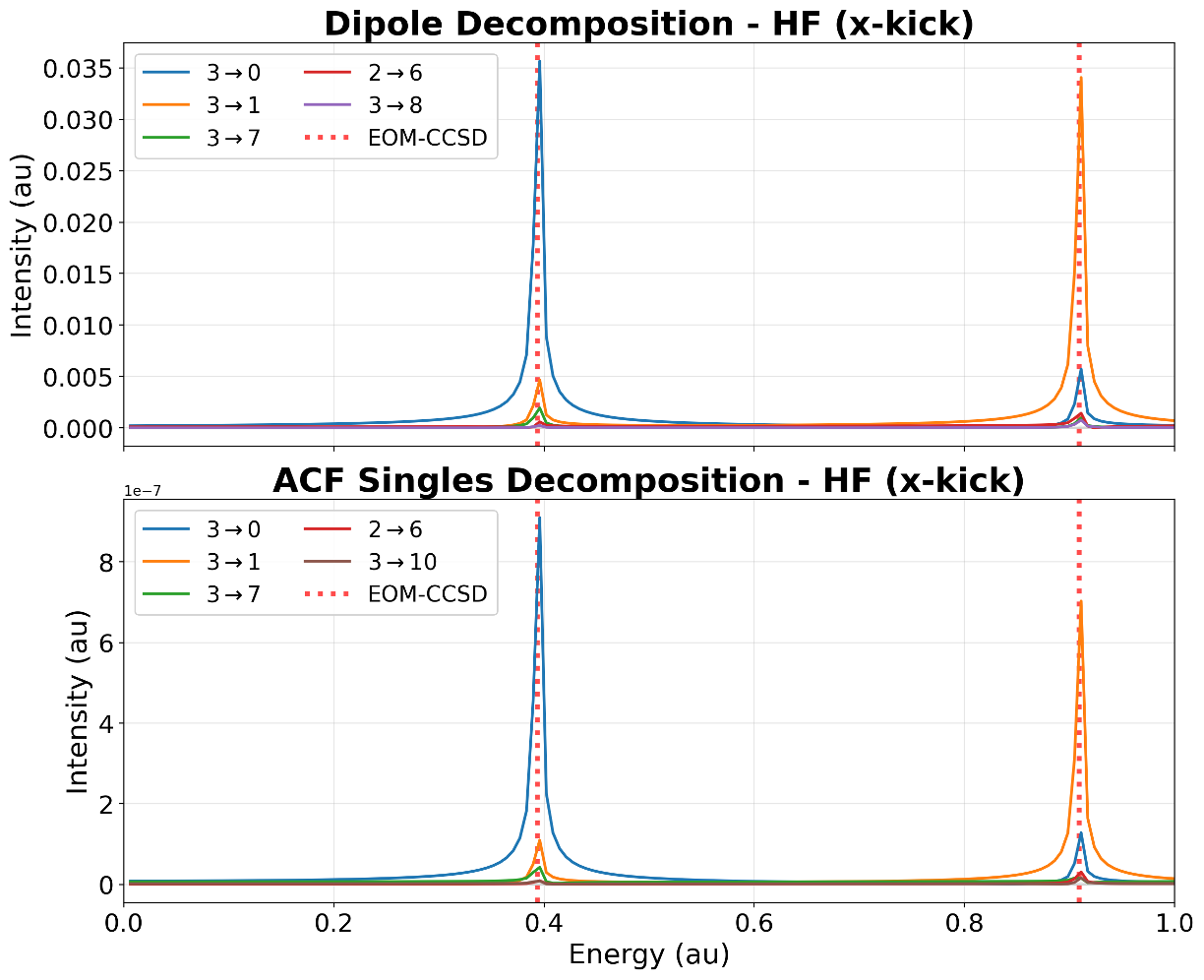}
    \caption{}
    \label{fig:hf_x}
  \end{subfigure}%
  \hfill
  \begin{subfigure}{0.5\textwidth}
    \centering
    \includegraphics[width=\textwidth]{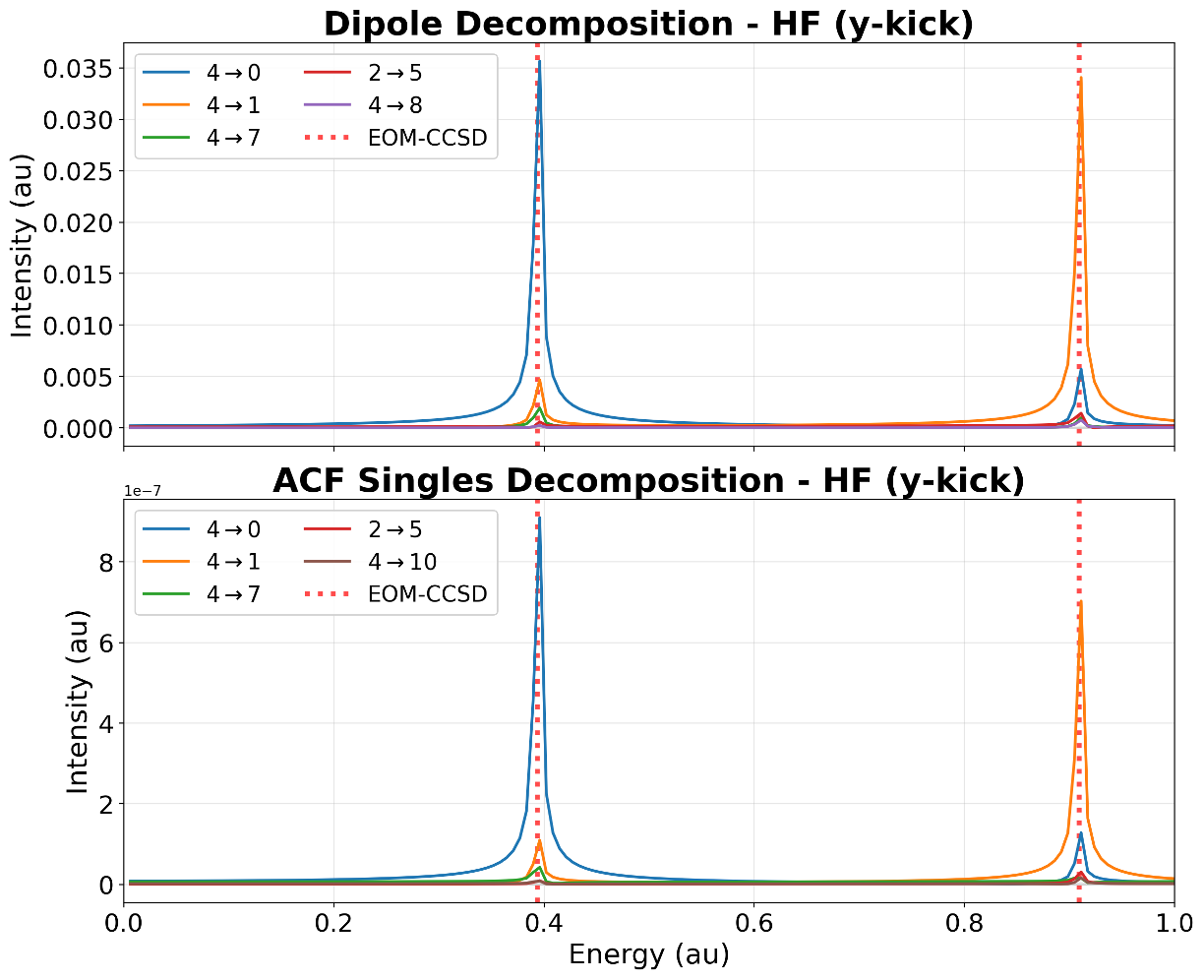}
    \caption{}
    \label{fig:hf_y}
  \end{subfigure}
  
  \vspace{10pt}
  
  \hspace*{\fill}%
  \begin{subfigure}{0.5\textwidth}
    \centering
    \includegraphics[width=\textwidth]{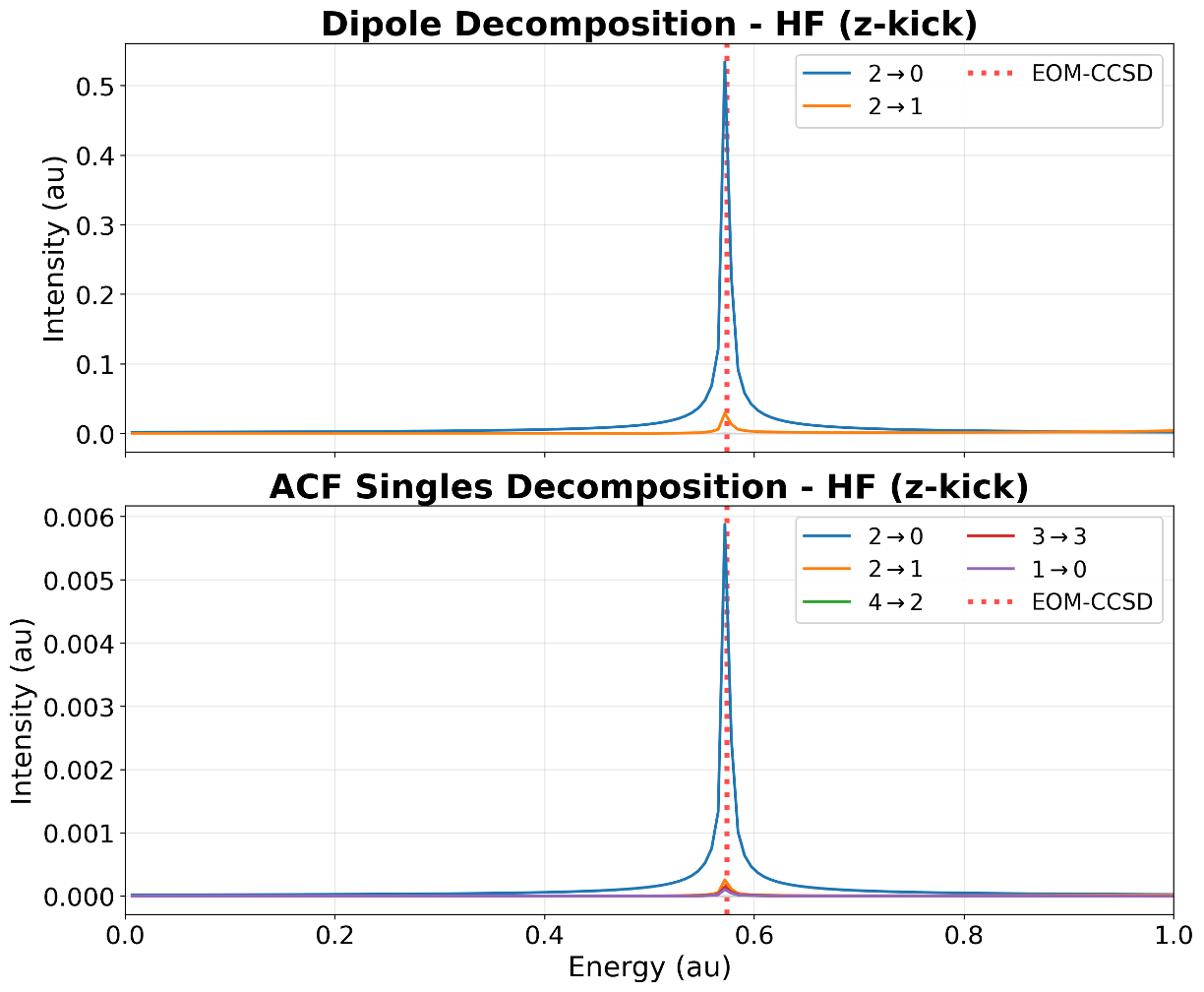}
    \caption{}
    \label{fig:hf_z}
  \end{subfigure}%
  \hspace*{\fill}%
  
  \captionsetup{skip=8pt}%
  \caption{Valence spectrum comparison of HF under (\subref{fig:hf_x}) $x$-polarized, (\subref{fig:hf_y}) $y$-polarized, and (\subref{fig:hf_z}) $z$-polarized perturbations at $0.01\,\mathrm{a.u.}$ field strength. Each panel shows dipole decomposition (top) and ACF singles decomposition (bottom). EOM-CCSD reference energies are indicated by red vertical dotted lines.}
  \label{fig:hf}
\end{figure}

Hydrogen fluoride, as the simplest heteronuclear diatomic molecule in this study, exhibits two prominent valence transitions at $0.394\,\mathrm{a.u.}$ and $0.909\,\mathrm{a.u.}$, corresponding to HOMO $\to$ LUMO and HOMO $\to$ LUMO+1 excitations, respectively.
Figure \ref{fig:hf} shows the valence spectral comparison for $x$-, $y$-, and $z$-polarized perturbations at $0.01\,\mathrm{a.u.}$ field strength.
While the molecule belongs to the $D_{\infty h}$ point group, our calculations are carried out in $C_{2v}$, with the bond axis oriented along the $z$-axis. There are five doubly occupied molecular orbitals (MOs), which we enumerate in energetic ordering from $0$ to $4$, are approximately described as: (0) F $1s$ core ($A_1)$; (1) $\sigma$ bonding MO ($A_1$); (2) $\sigma$ bonding MO with F $p_z$-type character ($A_1$); and (3) and (4) degenerate non-bonding $2p_x$- ($B_1$) and $2p_y$ ($B_2$) orbitals on F.  The virtual orbitals are enumerated separately from the occupied orbitals such that, e.g., $4 \to 0$ denotes the HOMO $\to$ LUMO orbital transition.  Virtual orbitals 0, 1, 2, 7, 8, and 13 are all of $\sigma$-type with varying levels of anti-bonding character; the rest occur as degenerate pairs.

The $x$- and $y$-polarized perturbations (perpendicular to the molecular axis) produce nearly identical spectra in both dipole and ACF decompositions (Figures \ref{fig:hf_x} and \ref{fig:hf_y}), arising from the degeneracy of the $p_x$ and $p_y$ orbitals on fluorine. For the $x$-polarized field, both methods show the $4 \to 0$ transition dominating the $0.394\,\mathrm{a.u.}$ peak and the $4 \to 1$ transition dominating the $0.909\,\mathrm{a.u.}$ peak.
The $y$-polarized fields show the same pattern with degenerate partner orbitals ($3 \to 0$ and $3 \to 1$), confirming that orbitals $3$ and $4$ represent the $p_x$ and $p_y$ components of the $\pi$ manifold.

In contrast, $z$-polarized perturbations (along the molecular axis) access a distinct subset of transitions below 1.0 a.u.\ (Figure \ref{fig:hf_z}), with a peak at $0.574\,\mathrm{a.u.}$ corresponding to the $2 \to 0$ transition. The perpendicular $\pi$ transitions that dominate $x$/$y$ spectra are naturally absent in the $z$ spectrum, while the axial $\sigma$ transitions are absent in perpendicular field directions. This directional selectivity validates that both decomposition methods correctly capture the symmetry-determined selection rules: perpendicular fields access $\pi \to \sigma^*$ transitions ($4 \to 0$, $3 \to 1$ from the $p_x$/$p_y$ orbitals) while axial fields access $\sigma \to \sigma^*$ transitions ($2 \to 0$, $2 \to 1$ from the $p_z$ orbital).

Table \ref{tab:hf} compares ACF decomposition with EOM-CCSD eigenvector analysis.
\begin{table}[h]
\centering
    \caption{Orbital character assignments from ACF decomposition at $\boldsymbol{0.01\,\mathrm{a.u.}}$ field strength and from EOM-CCSD for HF valence transitions.}
\label{tab:hf}
\begin{tabular}{lccccc}
\toprule
Energy & Symmetry & Field & Transition & ACF & EOM \\
(a.u.) &  & Polarization &  & amplitude & eigenvector \\
\midrule
0.394 & $\Pi$ & $x$ & 4 $\to$ 0 & 0.560 & 0.674 \\
      &       & $y$ & 3 $\to$ 0 & 0.560 & 0.674 \\
0.909 & $\Pi$ & $x$ & 4 $\to$ 1 & 0.554 & 0.669 \\
      &       & $y$ & 3 $\to$ 1 & 0.554 & 0.669 \\
\midrule
0.574 & $\Sigma$ & $z$ & 2 $\to$ 0 & 0.675 & 0.687 \\
\bottomrule
\end{tabular}
\end{table}
While these methods measure fundamentally different quantities---time-evolved cluster amplitudes versus static eigenvector components---normalization in the singles and doubles space (Eq. \eqref{eq:acf_n}) enables semi-quantitative comparison. Both methods consistently identify the same dominant orbital characters for all six transitions. ACF amplitudes of $\sim\!0.56$ for $\pi$ transitions compare well with EOM eigenvectors of $\sim\!0.67$. Critically, both methods show identical trends: degenerate orbital pairs ($3$ and $4$) yield identical amplitudes in both ACF ($0.560$) and EOM-CCSD ($0.674$), confirming proper treatment of $\Pi$ state degeneracy.

\subsubsection{Water}

\begin{figure}[htbp]
 \centering
 \vspace*{5pt}%
  \begin{subfigure}{0.5\textwidth}
    \centering
    \includegraphics[width=\textwidth]{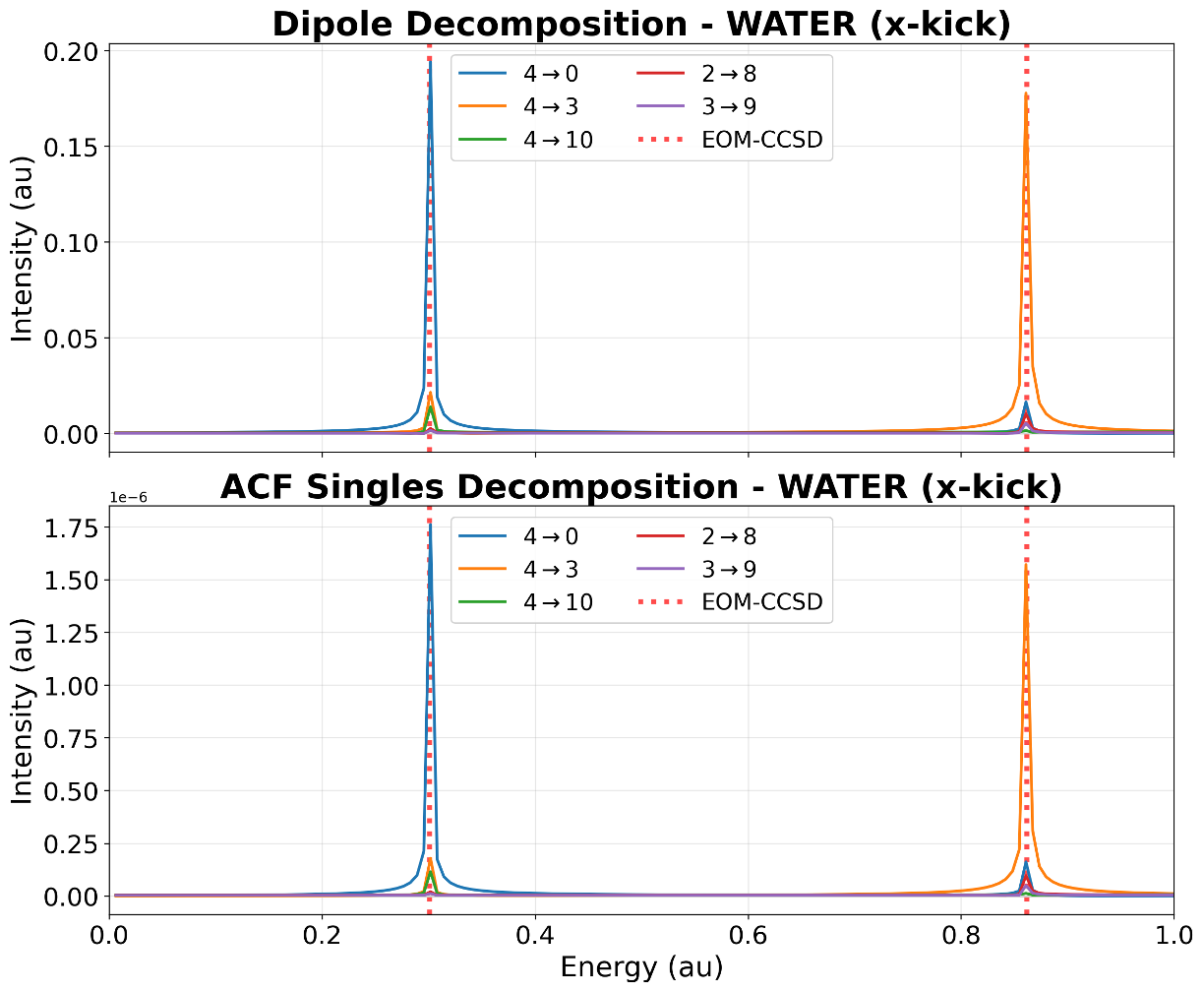}
    \caption{}
    \label{fig:h2o_x}
  \end{subfigure}%
  \hfill
  \begin{subfigure}{0.5\textwidth}
    \centering
    \includegraphics[width=\textwidth]{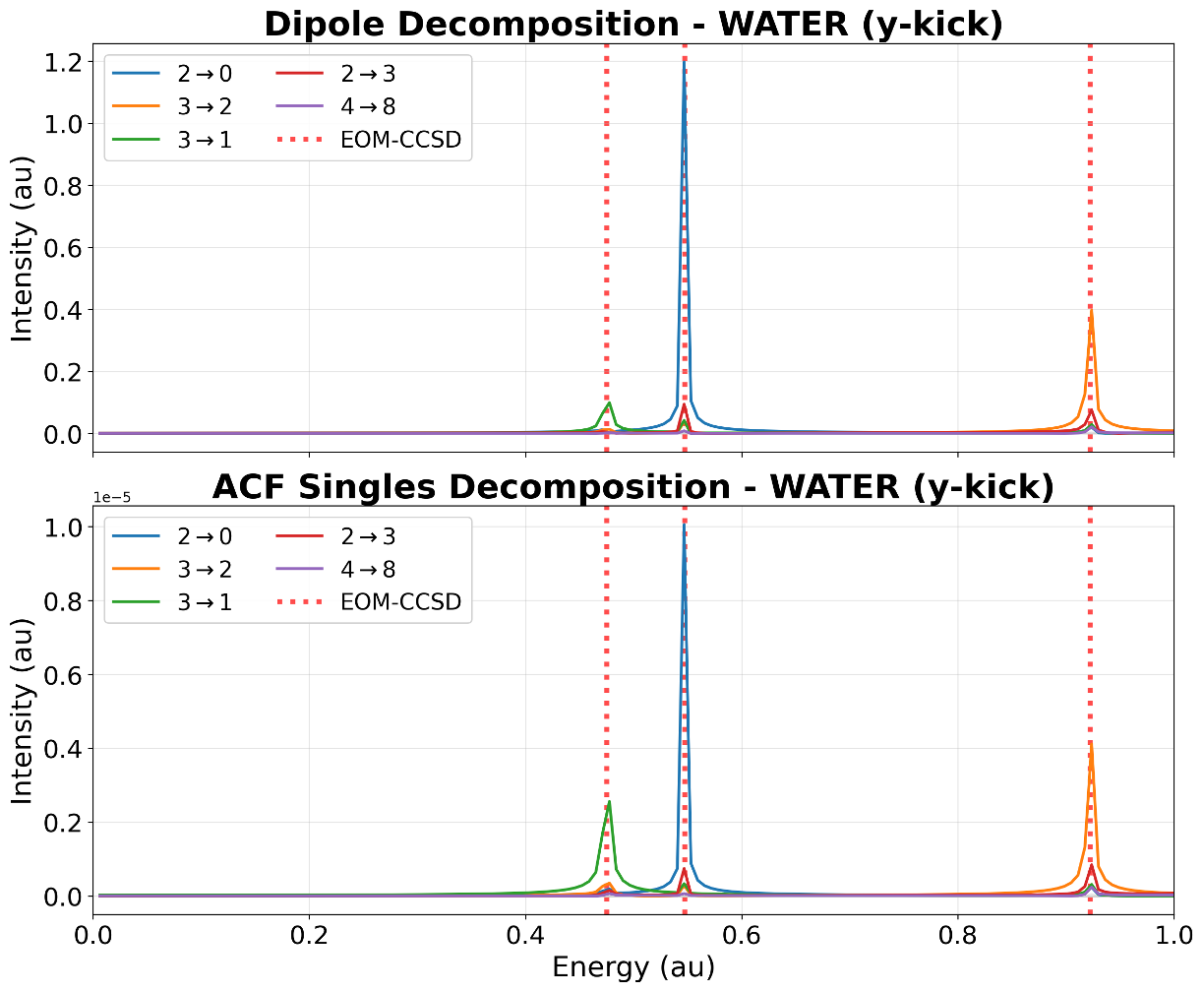}
    \caption{}
    \label{fig:h2o_y}
  \end{subfigure}
  
  \vspace{10pt}
  
  \hspace*{\fill}%
  \begin{subfigure}{0.5\textwidth}
    \centering
    \includegraphics[width=\textwidth]{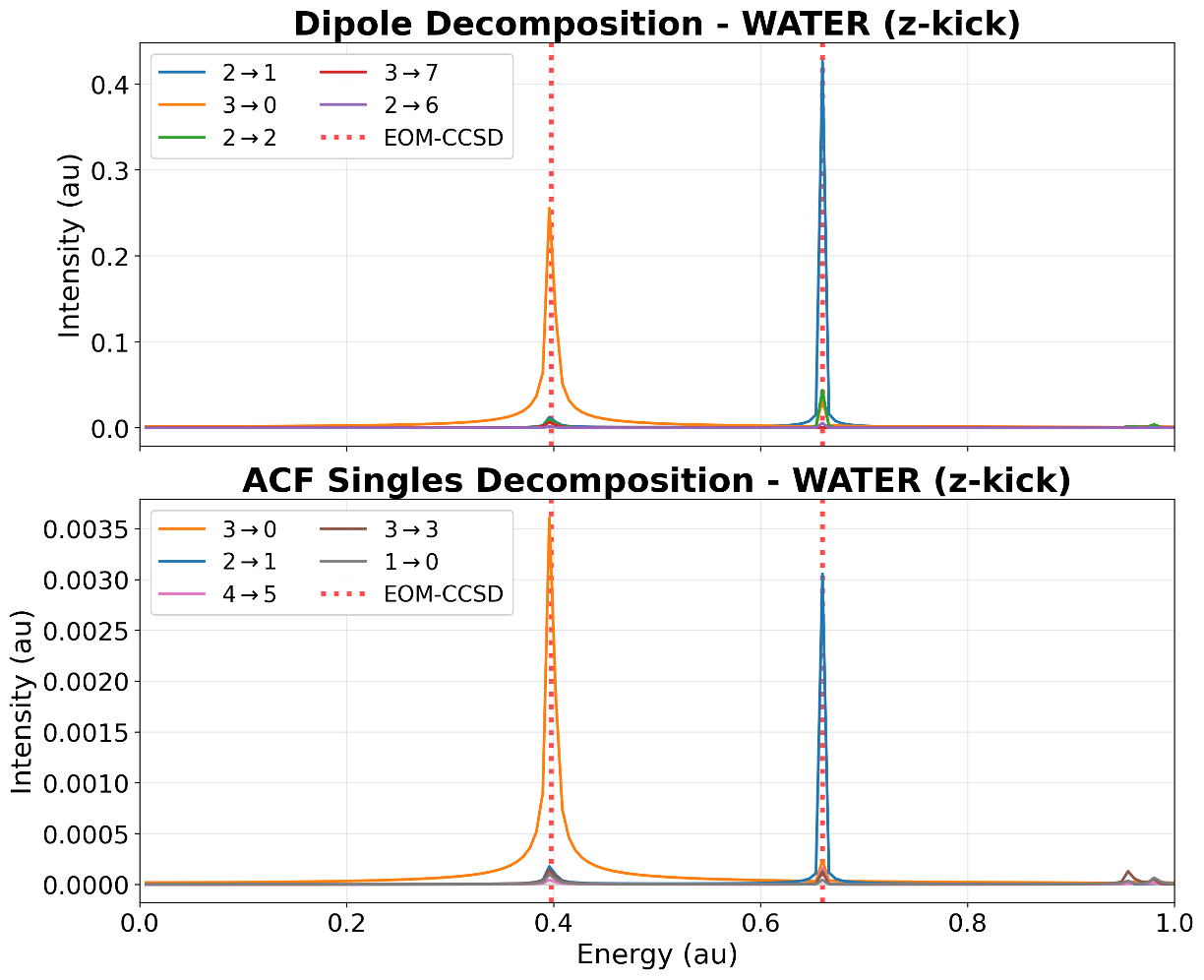}
    \caption{}
    \label{fig:h2o_z}
  \end{subfigure}%
  \hspace*{\fill}%
  
  \captionsetup{skip=8pt}%
    \caption{Valence spectrum comparison of \ce{H2O} under (\subref{fig:h2o_x}) $x$-polarized, (\subref{fig:h2o_y}) $y$-polarized, and (\subref{fig:h2o_z}) $z$-polarized perturbations at $0.01\,\mathrm{a.u.}$ field strength. Each panel shows dipole decomposition (top) and ACF singles decomposition (bottom). EOM-CCSD reference energies are indicated by red vertical dotted lines.}
\label{fig:h2o}
\end{figure}

Water exhibits strong directional selectivity in its valence excitation spectrum, with each Cartesian direction naturally accessing distinct subsets of transitions. Figure \ref{fig:h2o} shows the valence spectral comparison for $x$-, $y$-, and $z$-polarized perturbations at $0.01\,\mathrm{a.u.}$ field strength. 
For H$_2$O in C$_{2v}$ symmetry oriented in the $yz$-plane with its $C_2$ axis aligned along the $z$-axis (cf.\ Table S2 of the SI), the three highest-lying occupied orbitals are: MO $4$ (O $2p_x$ lone pair, $B_1$); MO $3$ (bonding MO involving O $2p_z$, $A_1$); and MO $2$ (bonding MO involving O $2p_y$, $B_2$). As for HF above, these transform as $x$, $z$, and $y$ coordinates respectively under C$_{2v}$ symmetry operations, which is relevant to the directional selectivity observed in the spectra.

The $x$-polarized field (Figure \ref{fig:h2o_x}) produces two dominant peaks at $0.300\,\mathrm{a.u.}$ and $0.861\,\mathrm{a.u.}$, corresponding to the $4 \to 0$ and $4 \to 3$ transitions. The $y$-polarized field (Figure \ref{fig:h2o_y}) accesses a completely different manifold with prominent peaks at $0.475\,\mathrm{a.u.}$, $0.547\,\mathrm{a.u.}$, and $0.922\,\mathrm{a.u.}$, corresponding to the $3 \to 1$, $2 \to 0$, and $3 \to 2$ transitions. The $z$-polarized field (Figure \ref{fig:h2o_z}) produces yet another distinct pattern with major peaks at $0.397\,\mathrm{a.u.}$ and $0.659\,\mathrm{a.u.}$, dominated by the $3 \to 0$ and $2 \to 1$ transitions, respectively. 

Both dipole and ACF decomposition methods yield consistent orbital assignments and relative intensities across all three field directions. However, ACF decomposition systematically reveals additional weak features at higher energies ($\sim\!1.0$--$1.5\,\mathrm{a.u.}$, not depicted) that are near or below the detection threshold in dipole spectra. By probing state populations directly rather than ground-state transition moments, ACF decomposition identifies transitions with small or zero oscillator strengths that would be missed by dipole-based analysis alone.

Table \ref{tab:h2o} compares ACF decomposition with EOM-CCSD eigenvector analysis across all three field directions.
\begin{table}[h]
\centering
    \caption{Orbital character assignments for \ce{H2O} from ACF decomposition at $\mathbf{0.01\,\mathrm{a.u.}}$ field strength and from EOM-CCSD.}
\label{tab:h2o}
\begin{tabular}{lccccc}
\toprule
Energy & Symmetry & Field & Transition & ACF & EOM \\
(a.u.) &  & Polarization &  & amplitude & eigenvector \\
\midrule
0.300 & B$_1$ & $x$ & 4 $\to$ 0 & 0.701 & 0.685 \\
0.861 & B$_1$ & $x$ & 4 $\to$ 3 & 0.702 & 0.676 \\
\midrule
0.475 & B$_2$ & $y$ & 3 $\to$ 1 & 0.697 & 0.682 \\
0.547 & B$_2$ & $y$ & 2 $\to$ 0 & 0.704 & 0.688 \\
0.922 & B$_2$ & $y$ & 3 $\to$ 2 & 0.688 & 0.662 \\
\midrule
0.397 & A$_1$ & $z$ & 3 $\to$ 0 & 0.672 & 0.682 \\
0.659 & A$_1$ & $z$ & 2 $\to$ 1 & 0.701 & 0.679 \\
\bottomrule
\end{tabular}
\end{table}
Both methods agree on the same dominant orbital characters for all seven transitions examined. The ACF amplitudes, normalized in the singles and doubles space, show consistent magnitudes across all transitions (ACF: $0.67$--$0.70$, EOM: $0.65$--$0.69$) with parallel trends. This internal consistency, combined with perfect agreement in orbital character assignments, validates that ACF decomposition correctly captures the electronic structure and symmetry-determined selection rules for molecules with C$_{2v}$ or lower symmetry.

\subsubsection{Ammonia}

\begin{figure}[htbp]
 \centering
 \vspace*{5pt}%
  \begin{subfigure}{0.5\textwidth}
    \centering
    \includegraphics[width=\textwidth]{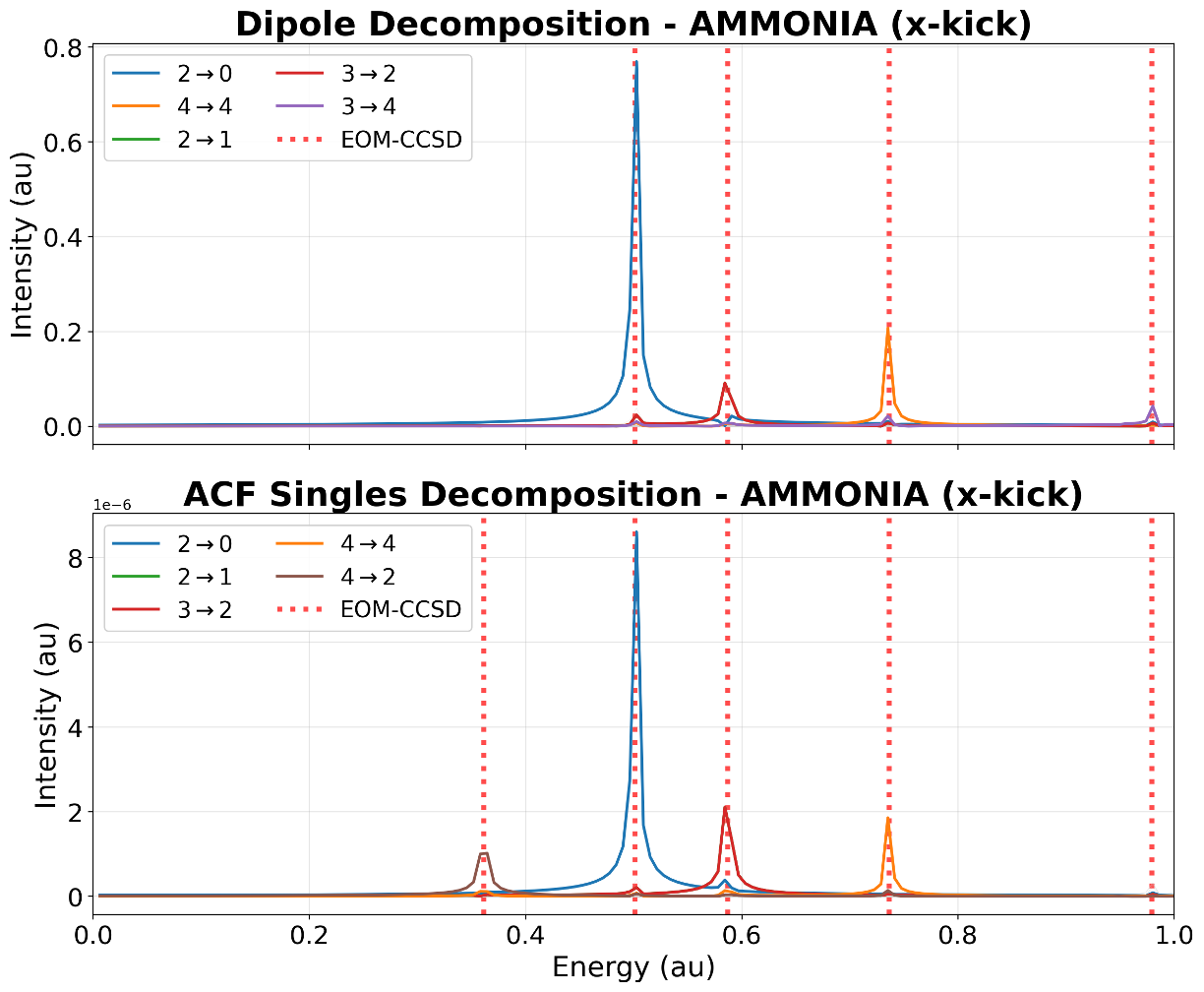}
    \caption{}
    \label{fig:nh3_x}
  \end{subfigure}%
  \hfill
  \begin{subfigure}{0.5\textwidth}
    \centering
    \includegraphics[width=\textwidth]{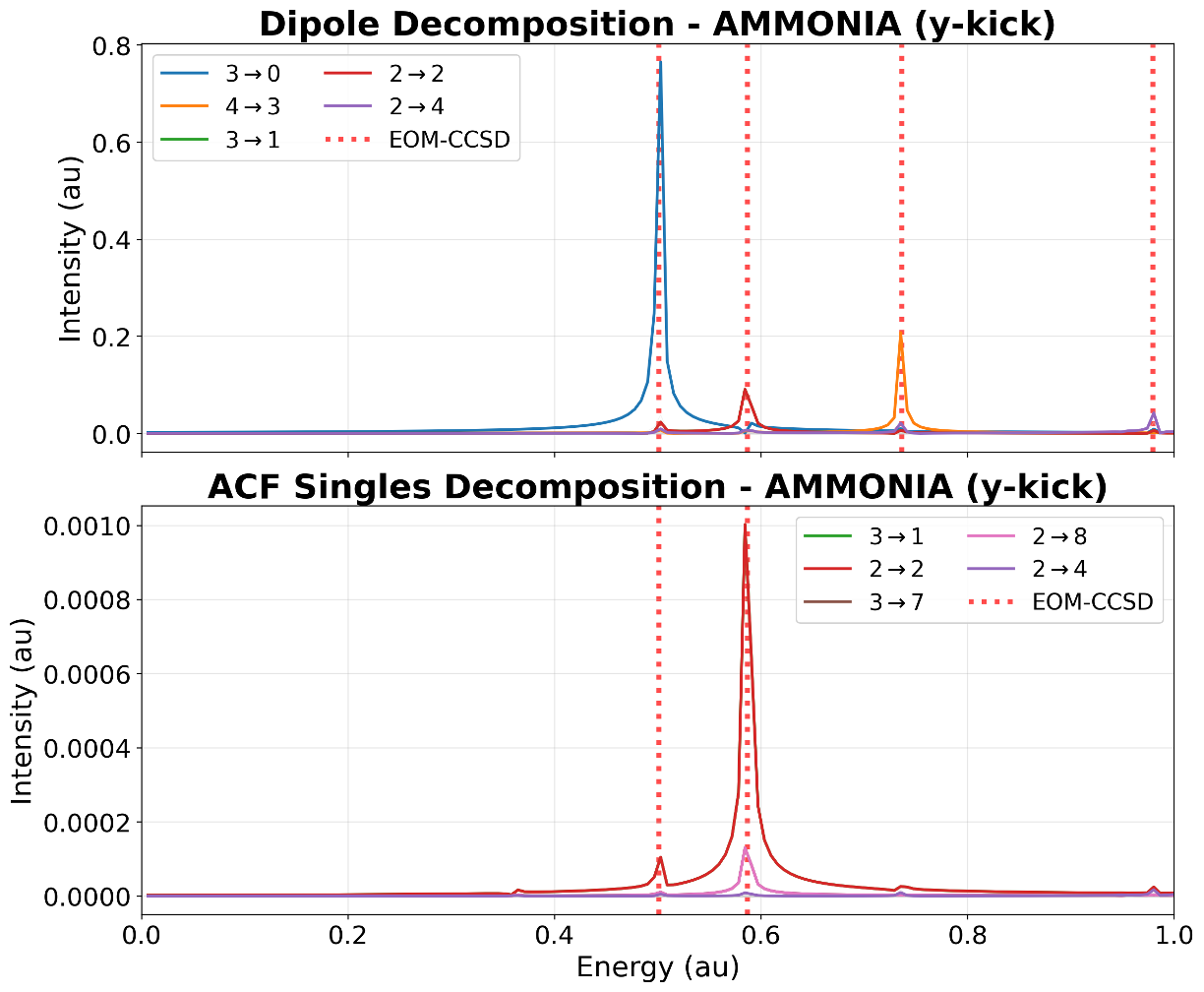}
    \caption{}
    \label{fig:nh3_y}
  \end{subfigure}
  
  \vspace{10pt}
  
  \hspace*{\fill}%
  \begin{subfigure}{0.5\textwidth}
    \centering
    \includegraphics[width=\textwidth]{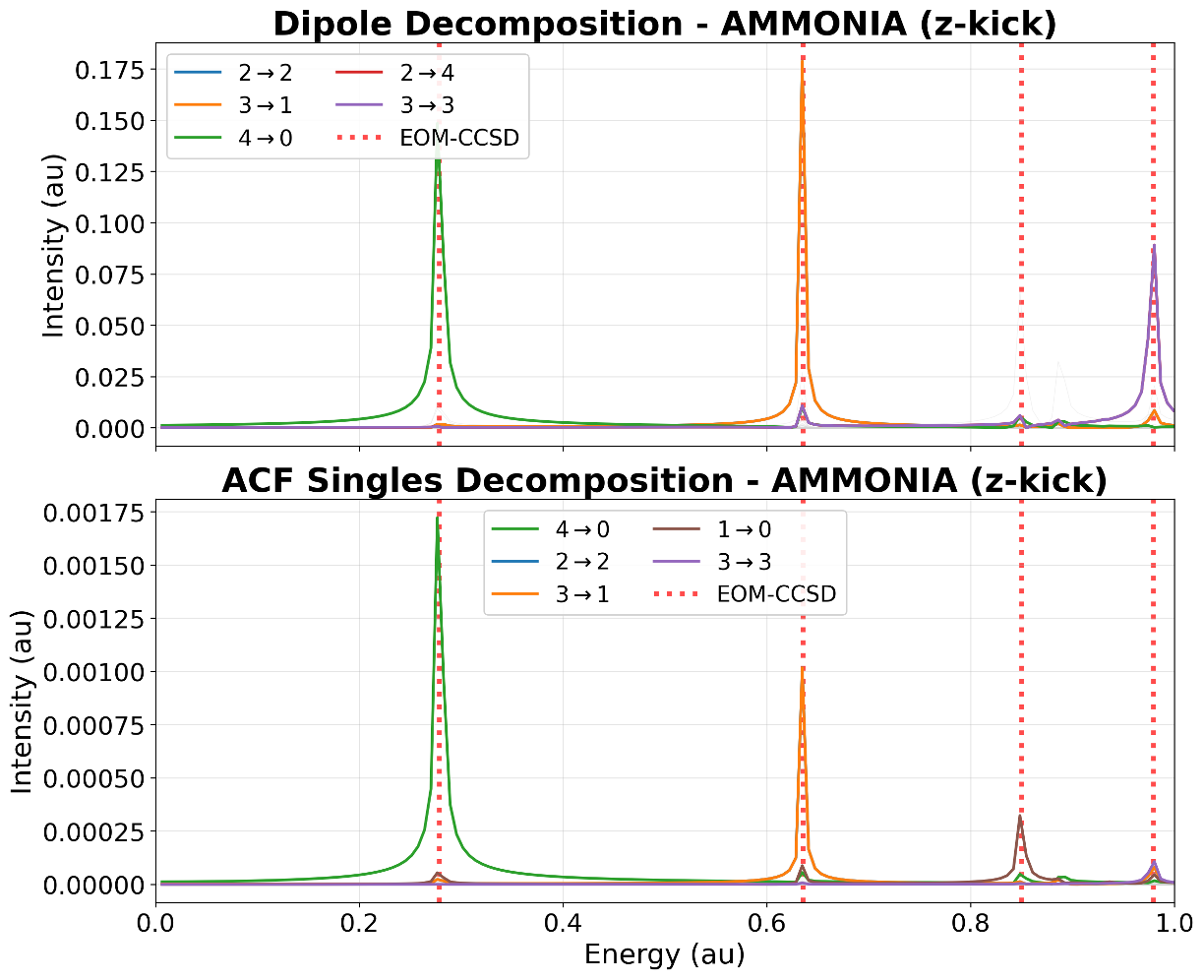}
    \caption{}
    \label{fig:nh3_z}
  \end{subfigure}%
  \hspace*{\fill}%
  
  \captionsetup{skip=8pt}%
    \caption{Valence spectrum comparison of \ce{NH3} with (\subref{fig:nh3_x}) $x$-polarized, (\subref{fig:nh3_y}) $y$-polarized, and (\subref{fig:nh3_z}) $z$-polarized perturbations at $0.01\,\mathrm{a.u.}$ field strength. Top panels show dipole decomposition, bottom panels show ACF singles decomposition. EOM-CCSD reference energies are indicated by red vertical dotted lines.}
\label{fig:nh3}
\end{figure}

Ammonia reveals an interesting phenomenon that demonstrates the somewhat complementary nature of the dipole and ACF decomposition methods.
Figure \ref{fig:nh3} shows the valence spectral comparison for $x$-, $y$-, and $z$-polarized perturbations at $0.01\,\mathrm{a.u.}$ field strength.  
Note that the molecular $C_3$ symmetry axis is aligned along the $z$-axis, and one of the N$-$H bonds is chosen to lie in the $yz$-plane.  (See the geometry provided in the SI.)  While the molecule belongs to the $C_{3v}$ point group, our calculations are limited to $C_s$, i.e.\ the largest Abelian subgroup of $C_{3v}$.  The three highest-energy occupied MOs in this case are the N 2$p_z$ lone pair (MO 4, $A'$) and the degenerate pair of N$-$H bonding MOs, one of which involves bonding with the N $2p_x$ orbital and thus has a node in the $yz$ plane (MO 3, $A''$) and the other involves bonding with the N $2p_y$ orbital (MO 2, $A'$).  We also note the presence of several important degenerate pairs of virtual MOs: 1 ($A'$) and 2 ($A''$), 3 ($A'$) and 4 ($A''$), 7 ($A'$) and 8 ($A''$) that will appear in the discussion below.

Focusing on the region of the spectrum below 1.0 a.u., the $z$-polarized field (Figure \ref{fig:nh3_z}) reveals significant transitions at 0.279, 0.635, 0.850 and 0.978 a.u., all of which correspond to non-degenerate states. The dipole and ACF decompositions agree on the primary MO contributions to the first, second, and fourth peaks, viz.\ $4\rightarrow 0$, $3\rightarrow 1$, and $3\rightarrow 3$, respectively. However, the third peak appears very weakly in the dipole decomposition, and, while the ACF decomposition for the peak at 0.850 a.u.\ indicates a major contribution from $1\rightarrow 0$, EOM-CCSD yields a larger contribution from $4\rightarrow 5$ (0.469), with $1\rightarrow 0$ somewhat smaller (0.227).

The $x$- and $y$-polarized fields (Figures \ref{fig:nh3_x} and \ref{fig:nh3_y}) produce visually identical dipole decomposition spectra, with major peaks below 1.0 a.u.\ at 0.501, 0.587, and 0.736 a.u.\ (all of which are doubly degenerate) showing the same intensities and peak shapes, as expected. This similarity reflects the fact that both $x$ and $y$ transform as components of the $E$ irreducible representation of $C_{3v}$ (though in our calculations in the $C_s$ subgroup the degeneracies are not exact). Furthermore, according to the dipole decomposition, the largest orbital contributions to these peaks all follow a similar pattern involving transitions to/from degenerate MOs.  For example, for the 0.501 a.u.\ peak, the $x$- and $y$-kicks reveal singles contributions into the non-degenerate virtual MO 0 from the degenerate MOs 2 and 3, respectively.  The particular pattern of MO contributions arises directly from the irreps of the MOs and the dipole-field components, e.g.\ the $2 (A'') \rightarrow 0 (A')$ transition couples through the $x (A'')$ component of the dipole moment while the $3 (A')\rightarrow 0 (A')$ transition coupled through the $y (A')$ component.  Similarly, the peak at 0.587 a.u.\ is characterized by transitions between degenerate occupied and virtual MOs: $2 (A'') \rightarrow 1 (A')$ and $3 (A') \rightarrow 2(A'')$ for the $x (A'')$ dipole component and $2 (A'') \rightarrow 2 (A'')$ and $3 (A') \rightarrow 1(A')$ for the $y (A')$ dipole component.  All of these transitions agree with the EOM-CCSD eigenvector components.  (Note that, for the 0.587 a.u.\ peak, only one of the two contributing MO transitions is visible in the plot because they lie exactly on top of each other.)

However, the $x$- and $y$-kick ACF decompositions behave very differently.  The $x$-kick ACF decomposition is visually very similar to the corresponding dipole decomposition, apart from a new peak at 0.361 a.u.\ corresponding to yet another degenerate state.  For this peak, the major MO contribution is $4 (A') \rightarrow 2 (A'')$, which are dipole-allowed, but also have a much smaller oscillator strength than the peaks at 0.501 and 0.586 a.u.  This is, therefore, an example of the ability of the ACF to reveal MO information on even weakly allowed states. The $y$-kick ACF decomposition, however, is strikingly different from both its dipole-decomposition and $x$-kick ACF counterparts.  For the large ACF peak at 0.587 a.u., for example, the primary MO transitions correspond to $2 (A'') \rightarrow 2 (A'')$ and $3 (A') \rightarrow 1(A')$ (though only one is visible in the plot because they are identical), which is in agreement with the $x$-kick ACF decomposition.  In addition, there is a visible contribution from the $2 (A'') \rightarrow 8 (A'')$ component that does not appear in either the dipole decomposition or among the top-five contributions to the EOM-CCSD eigenvector.  For the 0.501 a.u.\ peak, however, the same $2 \rightarrow 2$ and $3 \rightarrow 1$ contributions appear relatively strongly, even though they are much weaker in the corresponding dipole decomposition.

A clue to the source of these differences lies in a comparison of the intensities in the three ACF decomposition plots: the $x$ decomposition values are ca.\ two orders of magnitude smaller than those for $y$ and $z$. The reason for this can be determined by an analysis of the mathematical form of the ACF along with the symmetries of the MOs.  In the expression for the singles components of the ACF in Eq.~\eqref{eq:acf_a1}, the largest contributions arise from the $\lambda_a^i$ and $t_i^a$ components.  Indeed, an examination of the energy-domain values (obtained by Fourier transform of their imaginary components) of $\lambda_2^3$ and $t_3^2$  for the $x$-kick and of $\lambda_1^3$ and $t_3^1$ for the $y$-kick reveals \textit{identical} peaks at each of the transitions discussed above, with the largest peaks (ca.\ 0.35 in magnitude for $\lambda_2^3$/$\lambda_1^3$) centered at 0.587 a.u.  The difference in the ACF decomposition behavior lies in the fact that the ACF is obtained from the CC amplitudes starting at $t'$ when the field is first switched off.  (In this work, the delta pulse is applied at 0.01 a.u., so $t'=0.02$ a.u.)  For the $y$-kick, $\lambda_1^3 \approx 0.005$ and $t_3^1 \approx 0.003$ both before and immediately after the pulse because the field is relatively weak.  The subsequent oscillations of these and all the other amplitudes in time are what give rise to the peaks in the ACF (as well as the dipole decomposition and the absorption spectrum itself).  

For the $x$-kick, however, both $\lambda_2^3$ and $t_3^2$ are identically zero at $t=0.0$ because the MOs belong to different irreps (and are thus forbidden by symmetry).  After the field is switched off, they are non-zero, but still very small with values of ca.\ $1.2 \times 10^{-7}$ and $6.7 \times 10^{-8}$, respectively, because the application of the (weak) $x$-kick field breaks the $C_s$ mirror plane.  As a result, the $\lambda_a^i(t') t_i^a(t)$ and $\lambda_a^i(t) t_i^a(t')$ products appearing in the ACF singles decomposition in Eq.~\eqref{eq:acf_a1} differ by orders of magnitude between the $x$- and $y$-kicks. Thus, while the energy dependence of the $\lambda_a^i$ and $t_i^a$ amplitudes are identical for both $x$- and $y$-kicks, leading to the same energy-domain peaks, the fact that their initial values at $t'=0.02$ are so different in magnitude between the two kicks leads to the observed differences in the corresponding energy-domain $A_i^a(t',t)$ ACFs. (This would also be true for the doubles decomposition, $A_{ij}^{ab}(t',t)$ in Eq.~\eqref{eq:acf_a2}, but not for the $A0(t',t)$ in Eq.~\eqref{eq:acf_a0}.) Note also that the same observations hold identically for the $2\rightarrow 1$ and $2\rightarrow 2$ transitions for the $x$- and $y$-kick decompositions for the 0.587 a.u.\ peak.

\begin{figure}[htbp]
 \centering
 \includegraphics[width=\textwidth]{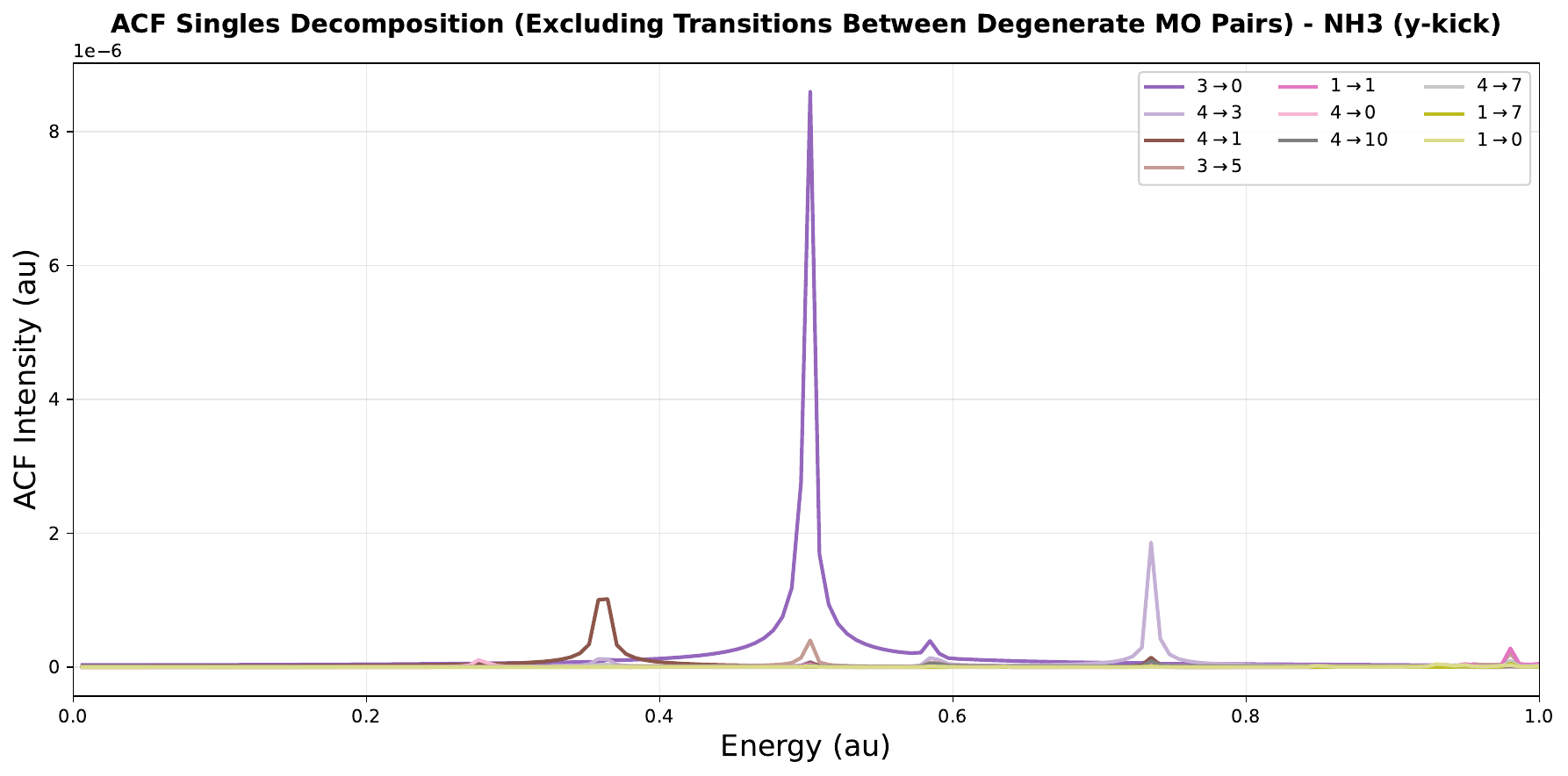}
 \caption{ACF singles decomposition of valence spectrum of \ce{NH3} with a $y$-polarized field, excluding transitions between the degenerate occupied MOs and degenerate pairs of virtual MOs.}
 \label{fig:nh3_nodeg}
\end{figure}

It is noteworthy that the large ACF decomposition contributions described above all are associated with transitions between degenerate pairs of both occupied and virtual MOs.  Figure \ref{fig:nh3_nodeg} depicts the NH$_3$ ACF singles decomposition for the $y$-kick field excluding such transitions, yielding intensity patterns nearly identical to those of the $x$-kick in Figure \ref{fig:nh3_x}.  Indeed, all the same peaks appear and with the same intensities, but the orbital contributions now refer to the $y$-kick partners of each degenerate MO pair.  For example, the largest peak at 0.501 a.u.\ has $3\rightarrow 0$ as its main contributor in the $y$-kick ACF, which corresponds to $2\rightarrow 0$ in the $x$-kick ACF.  In addition, while the $x$-kick ACF decomposition includes $2\rightarrow 1$ and $3\rightarrow 2$ transitions (as described above), the corresponding transitions of $2\rightarrow 2$ and $3\rightarrow 1$ in the $y$-kick ACF were masking the remaining components, e.g.\ in the 0.587 a.u.\ peak.  

\subsubsection{Methane}

\begin{figure}[htbp]
 \centering
 \vspace*{5pt}%
  \begin{subfigure}{0.5\textwidth}
    \centering
    \includegraphics[width=\textwidth]{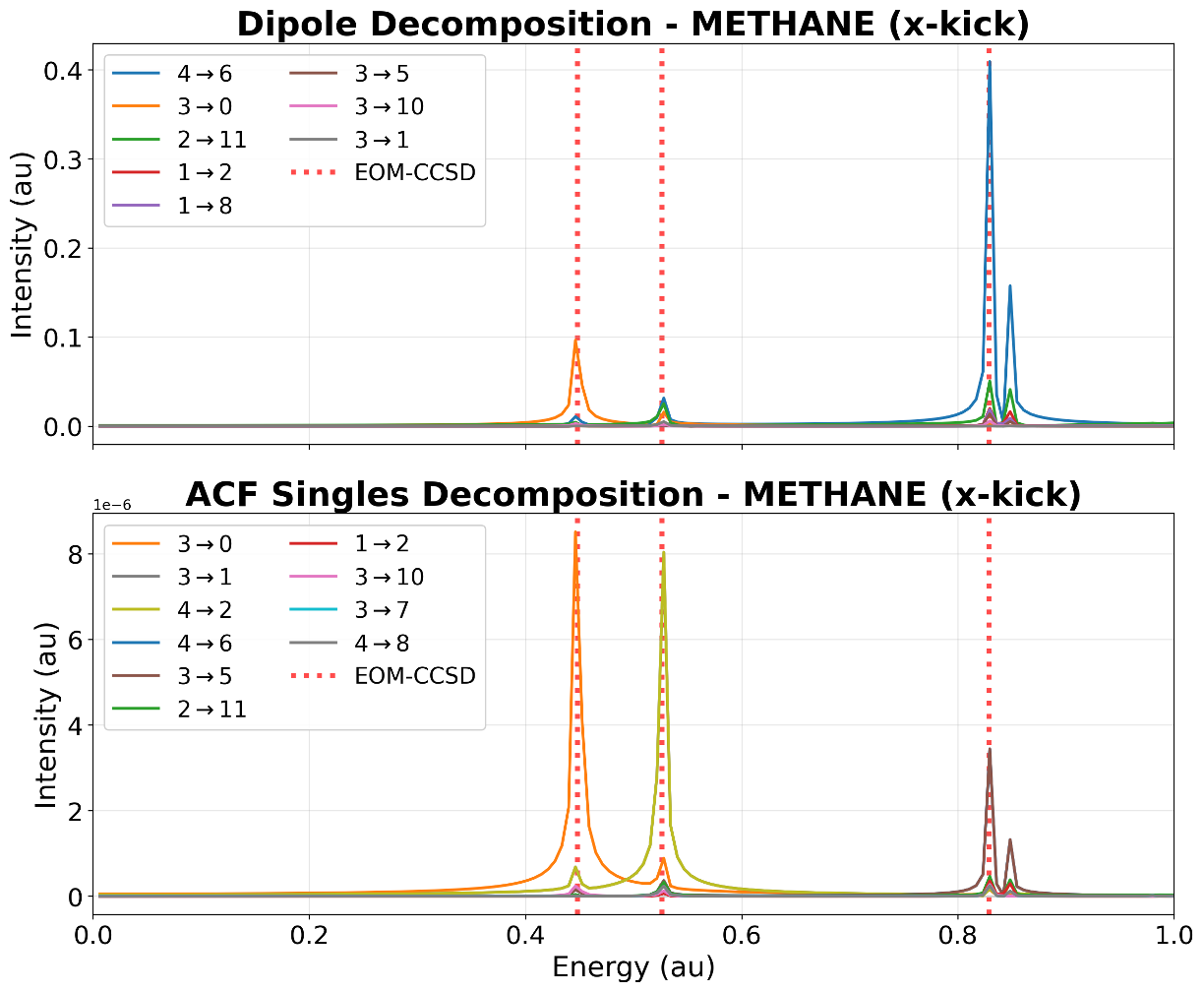}
    \caption{}
    \label{fig:ch4_x}
  \end{subfigure}%
  \hfill
  \begin{subfigure}{0.5\textwidth}
    \centering
    \includegraphics[width=\textwidth]{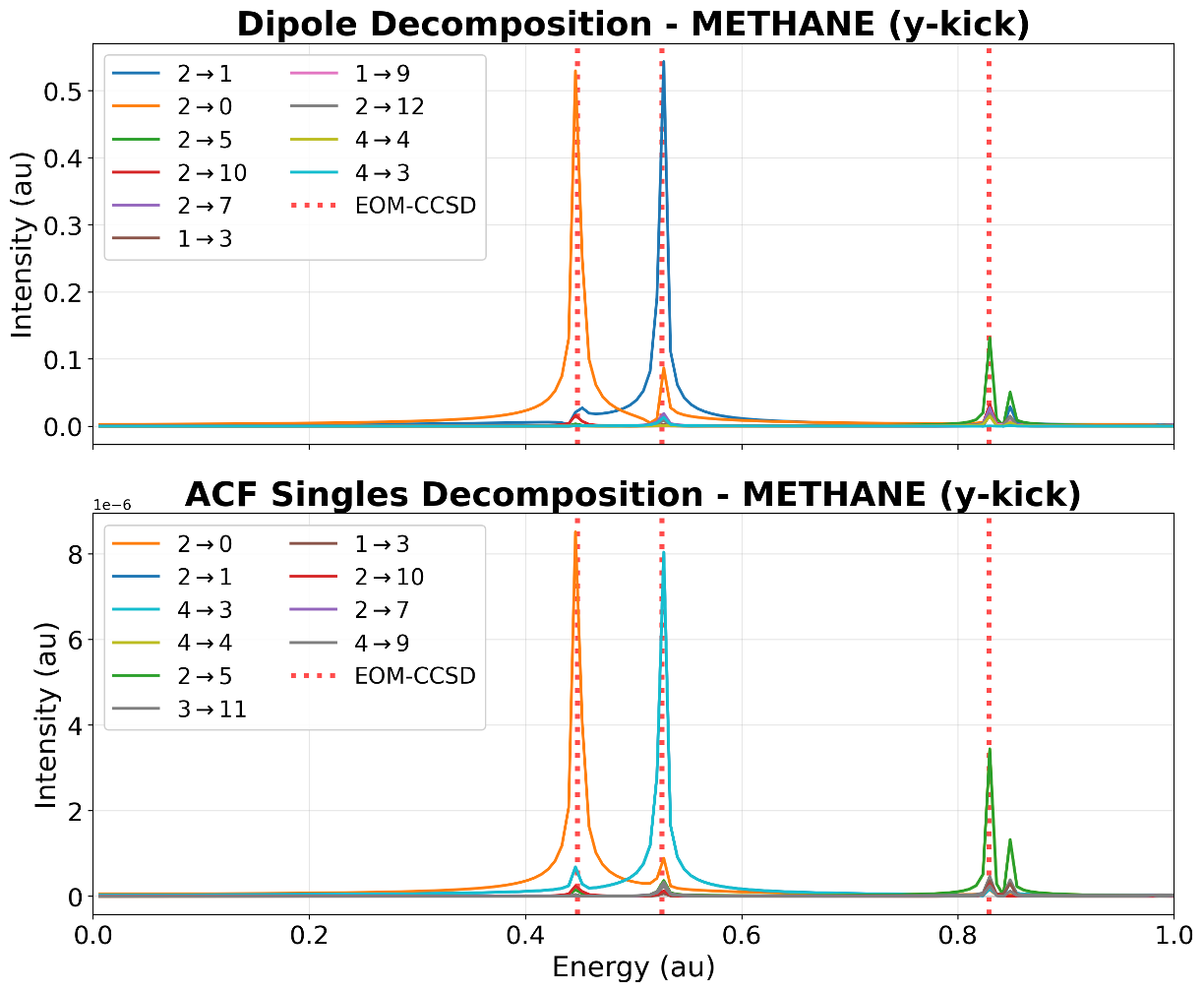}
    \caption{}
    \label{fig:ch4_y}
  \end{subfigure}
  
  \vspace{10pt}
  
  \hspace*{\fill}%
  \begin{subfigure}{0.5\textwidth}
    \centering
    \includegraphics[width=\textwidth]{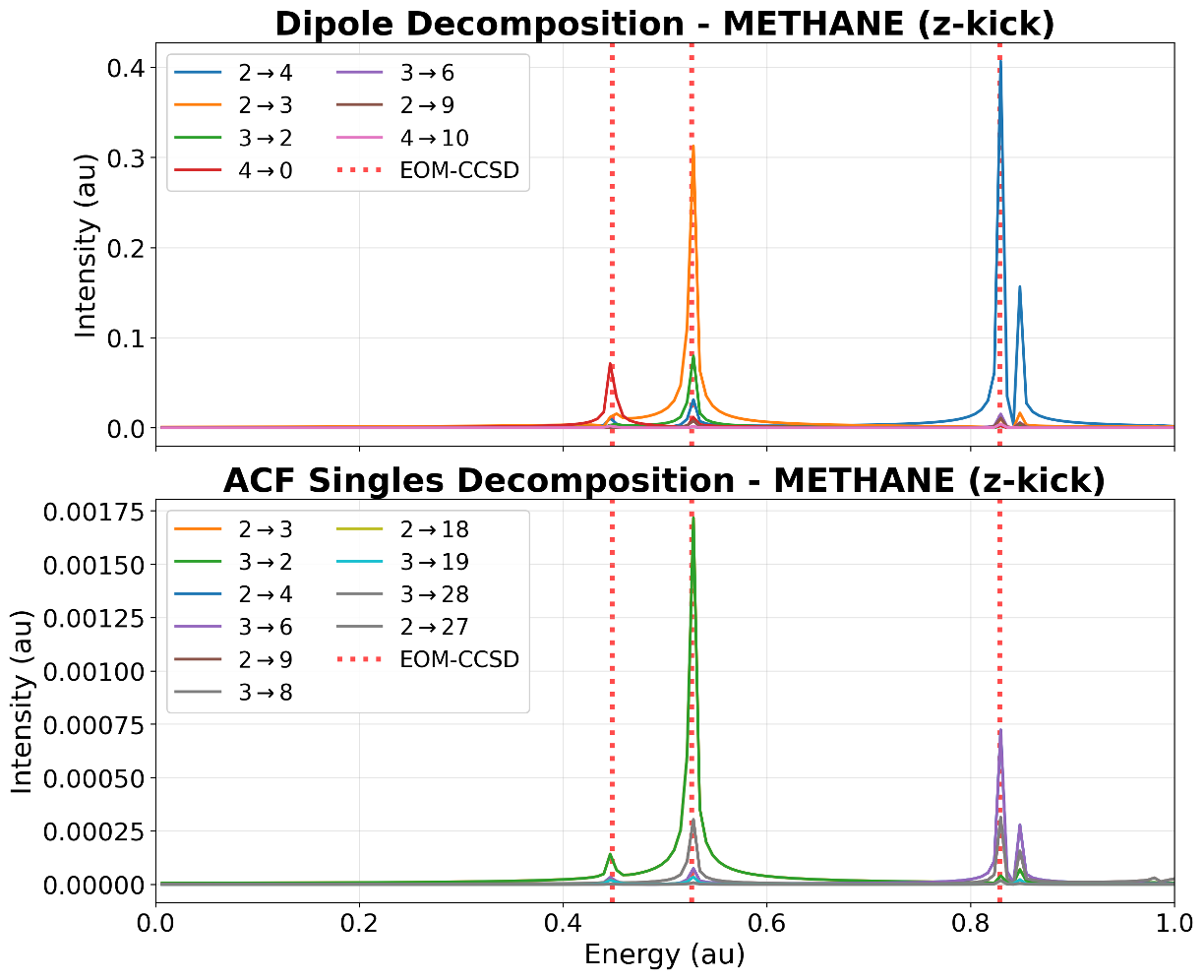}
    \caption{}
    \label{fig:ch4_z}
  \end{subfigure}%
  \hspace*{\fill}%
  
  \captionsetup{skip=8pt}%
    \caption{Full spectral comparison of \ce{CH4} under (\subref{fig:ch4_x}) $x$-polarized, (\subref{fig:ch4_y}) $y$-polarized, and (\subref{fig:ch4_z}) $z$-polarized perturbations at $0.01\,\mathrm{a.u.}$ field strength. Each panel shows dipole decomposition (top) and ACF singles decomposition (bottom). EOM-CCSD reference energies are indicated by red vertical lines.}
\label{fig:ch4}
\end{figure}

Methane presents the highest symmetry case in our study, with tetrahedral ($T_d$) point group symmetry producing triply degenerate $T_2$ states, and Figure \ref{fig:ch4} shows spectral comparisons for $x$-, $y$-, and $z$-polarized perturbations at $0.01\,\mathrm{a.u.}$ field strength.
We examine three valence transitions of $T_2$ symmetry at $0.448\,\mathrm{a.u.}$, $0.526\,\mathrm{a.u.}$, and $0.828\,\mathrm{a.u.}$, though our computations have been carried out in the $C_{2v}$ subgroup of $T_d$, giving the corresponding $A_1$, $B_1$, and $B_2$ states.  The underlying triply degenerate occupied MOs correspond to $B_2$ (MO 2), $B_1$ (MO 3), and $A_1$ (MO 4) irreps, while the non-degenerate valence occupied orbital (MO 1) transforms as the $A_1$ irrep.  Similarly, the lowest lying virtual MOs transform as follows: MO 0 ($A_1$), MOs 1-3 (triply degenerate, $A_1$, $B_1$, and $B_2$, respectively), and MOs 4-6 (triply degenerate, $B_2$, $A_1$, and $B_1$, respectively).   

The peak at 0.448 a.u.\ consists primarily of HOMO$\rightarrow$LUMO transitions from the triply-degenerate occupied MOs to the non-degenerate $A_1$ MO, according to the corresponding EOM-CCSD eigenvectors.  These are clearly represented in both the $x$- and $y$-kick dipole and ACF decompositions.  For the 0.526 a.u.\ peak, which is of mostly HOMO$\rightarrow$LUMO$+1$ character, the ACF decomposition agrees well with EOM-CCSD for the $x$- ($B_1$) and $y$-kick ($B_2$) fields, where the principal MO contributions are $3 (B_1)\rightarrow 1(A_1)/4 (A_1)\rightarrow 2 (B_1)$ and $4 (A_1)\rightarrow 3(B_2)/2 (B_2)\rightarrow 1 (A_1)$, respectively. (Note that only one of each pair is clearly visible in the Figure because the components are identical within each kick.)  The dipole decompositions for the 0.526 a.u.\ peak report not only these but also significant contributions from other transitions, e.g., the $2 (B_2)\rightarrow 11 (A_2)$ component for the $x$-kick.  Finally, the peak at 0.828 a.u.\ has mostly HOMO$\rightarrow$LUMO$+2$ character, which is confirmed by both the $x$- and $y$-kick ACF and dipole decompositions.  

\begin{figure}[htbp]
 \centering
 \includegraphics[width=\textwidth]{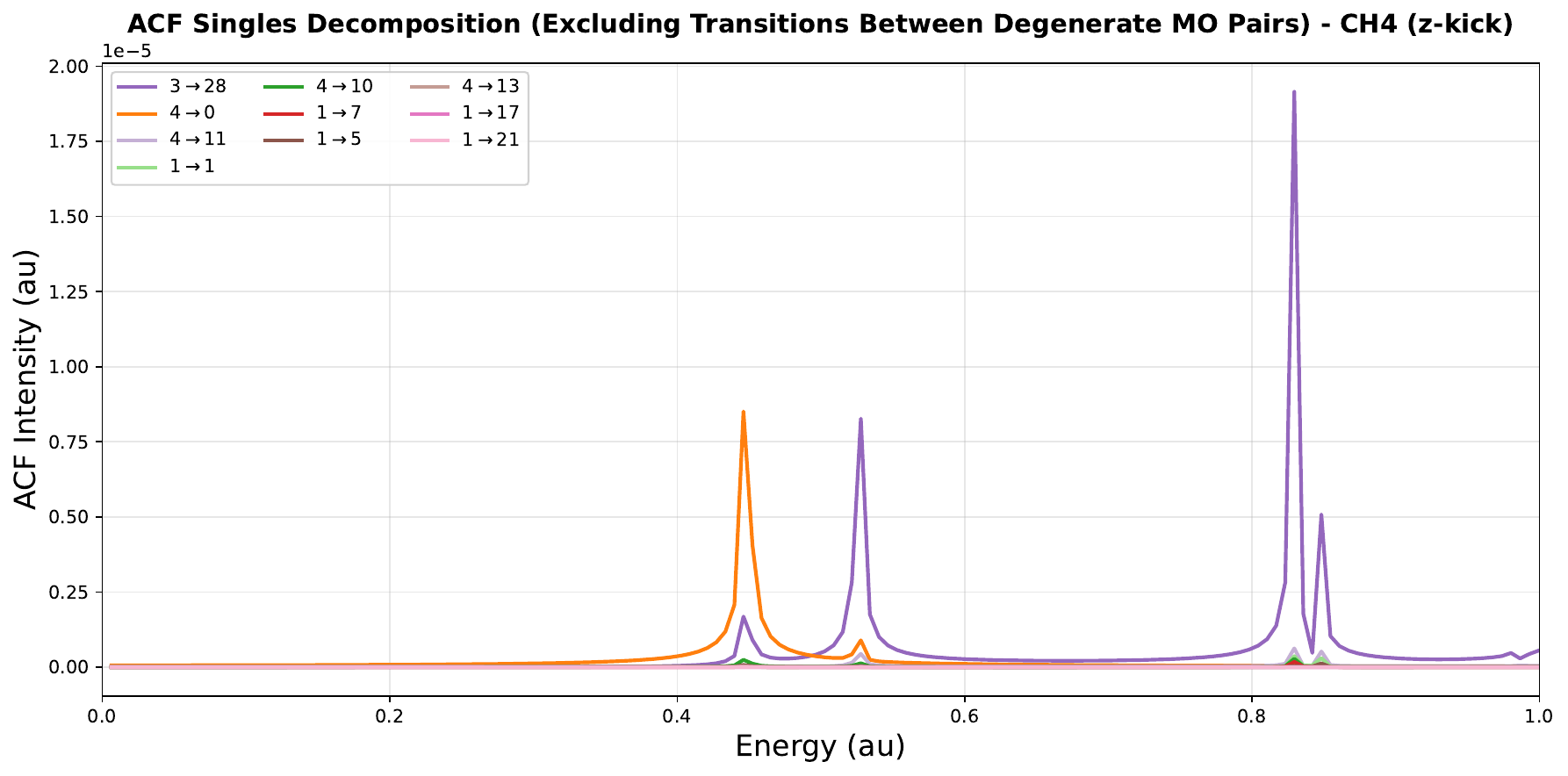}
 \caption{ACF singles decomposition of valence spectrum of \ce{CH4} with a $z$-polarized field, excluding transitions between the degenerate occupied MOs and degenerate pairs of virtual MOs.}
 \label{fig:ch4_nodeg}
\end{figure}

The $z$-kick field, the dipole decomposition identifies the corresponding $A_1$ components for each peak as its $x$- and $y$-kick counterparts, in agreement with EOM-CCSD.  The ACF decomposition, however, produces much larger values for each peak with the $z$-kick than for $x$- and $y$-kick, similar to the observations above for NH$_3$ with the $y$-kick field.  As a result, the $2 (B_1)\rightarrow 3 (B_1)$ and $3 (B_2)\rightarrow 2 (B_2)$ contributions to the 0.526 a.u.\ peak greatly dominate the region of the spectrum around 0.5 a.u., thus hiding the $4 (A_1)\rightarrow 0 (A_1)$ contribution to the 0.448 a.u.\ peak.  (The $2 (B_2)\rightarrow 4 (B_2)$ and $3 (B_1)\rightarrow 6 (B_1)$ contributions to the 0.828 a.u.\ peak are still preserved, though only one of these identical peaks is visible in the Figure.)  The $z$-kick ACF decomposition for CH$_4$ excluding transitions between degenerate sets of MO is shown in Figure~\ref{fig:ch4_nodeg}.  This reveals the  contribution from the $4 (A_1)\rightarrow 0 (A_1)$ contribution to the 0.448 a.u.\ peak, which is the largest in the EOM-CCSD eigenvector, along with other higher-energy MO contribution to the 0.525 and 0.828 a.u.\ peaks that are not significant according to EOM-CCSD.

We note that all orbital assignments reported here are basis-dependent, reflecting the canonical HF orbital representation. Unitary transformations within the occupied or virtual manifolds would redistribute ACF and dipole amplitudes among different orbital pairs while preserving total spectra and eigenstate characters. For high-symmetry systems like \ce{CH4}, a symmetry-adapted orbital basis (e.g., explicitly constructing $T_2$ linear combinations) might provide more direct physical interpretation than canonical orbitals. However, canonical orbitals offer computational simplicity and facilitate systematic comparison across molecules with different symmetries, which motivated our choice throughout this study.

\begin{figure}[htbp]
 \centering
 \vspace*{5pt}%
  \begin{subfigure}{0.5\textwidth}
    \centering
    \includegraphics[width=\textwidth]{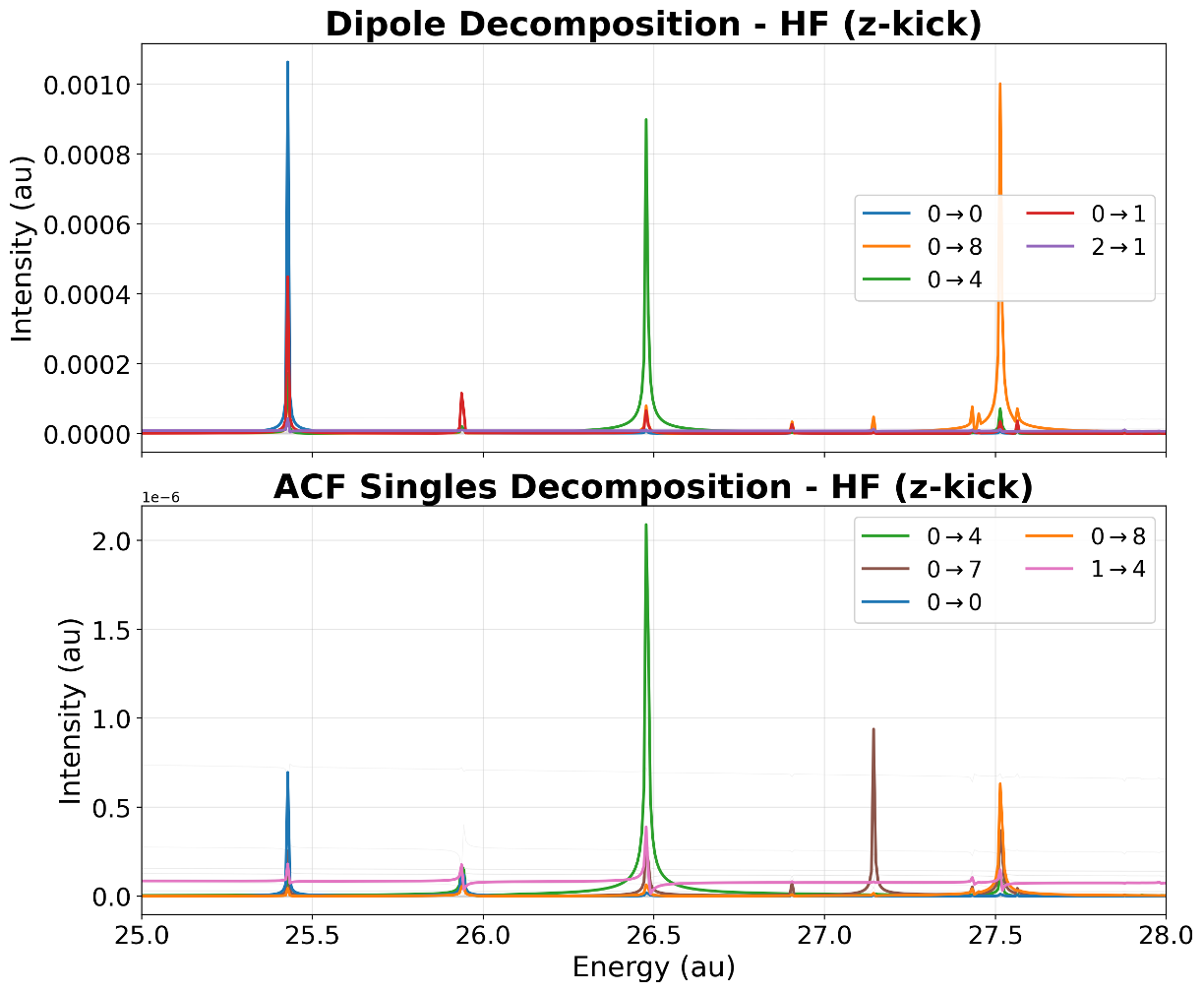}
    \caption{}
    \label{fig:hf_core}
  \end{subfigure}%
  \hfill
  \begin{subfigure}{0.5\textwidth}
    \centering
    \includegraphics[width=\textwidth]{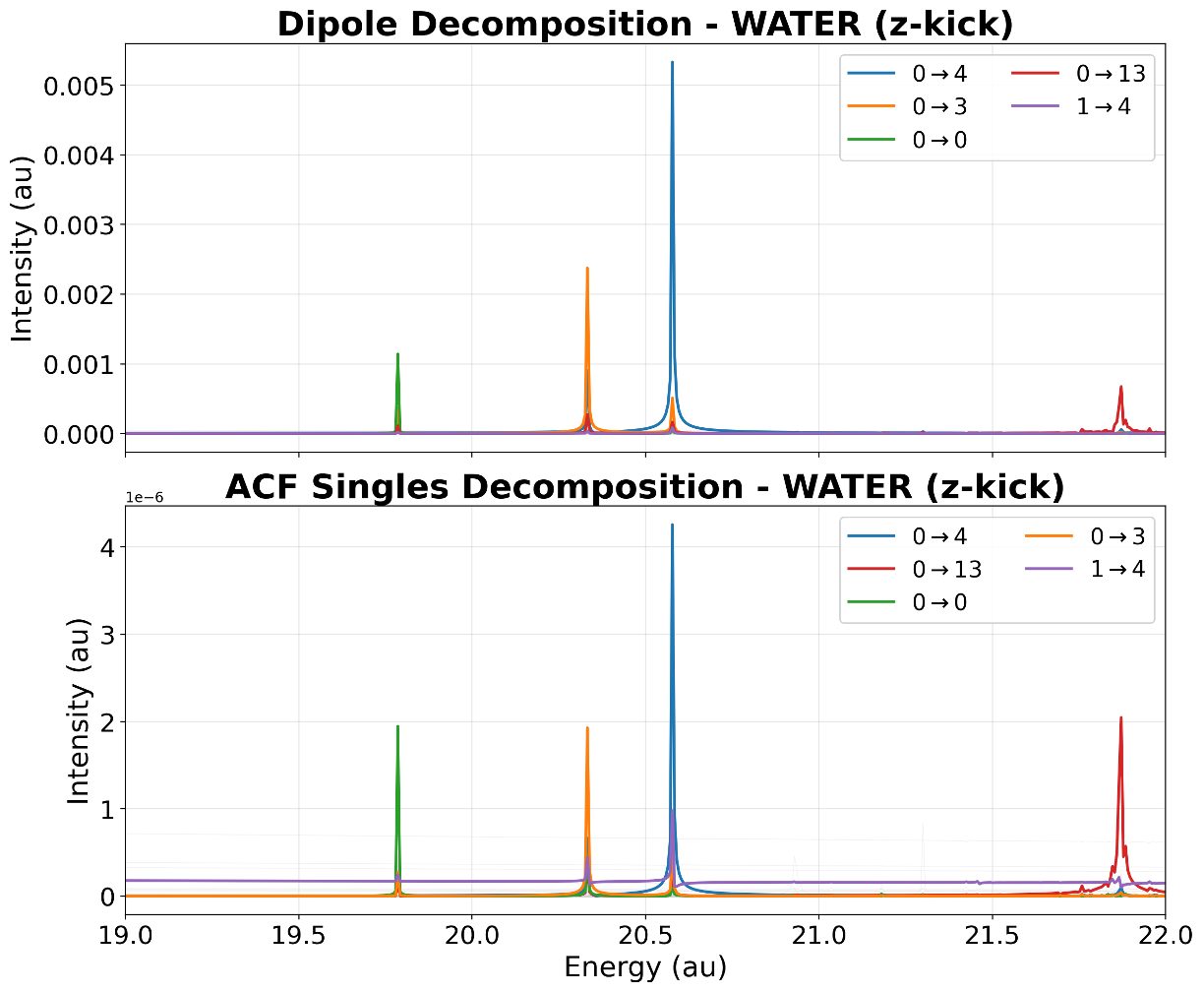}
    \caption{}
    \label{fig:water_core}
  \end{subfigure}
  
  \vspace{10pt}
  
  \hspace*{\fill}%
  \begin{subfigure}{0.5\textwidth}
    \centering
    \includegraphics[width=\textwidth]{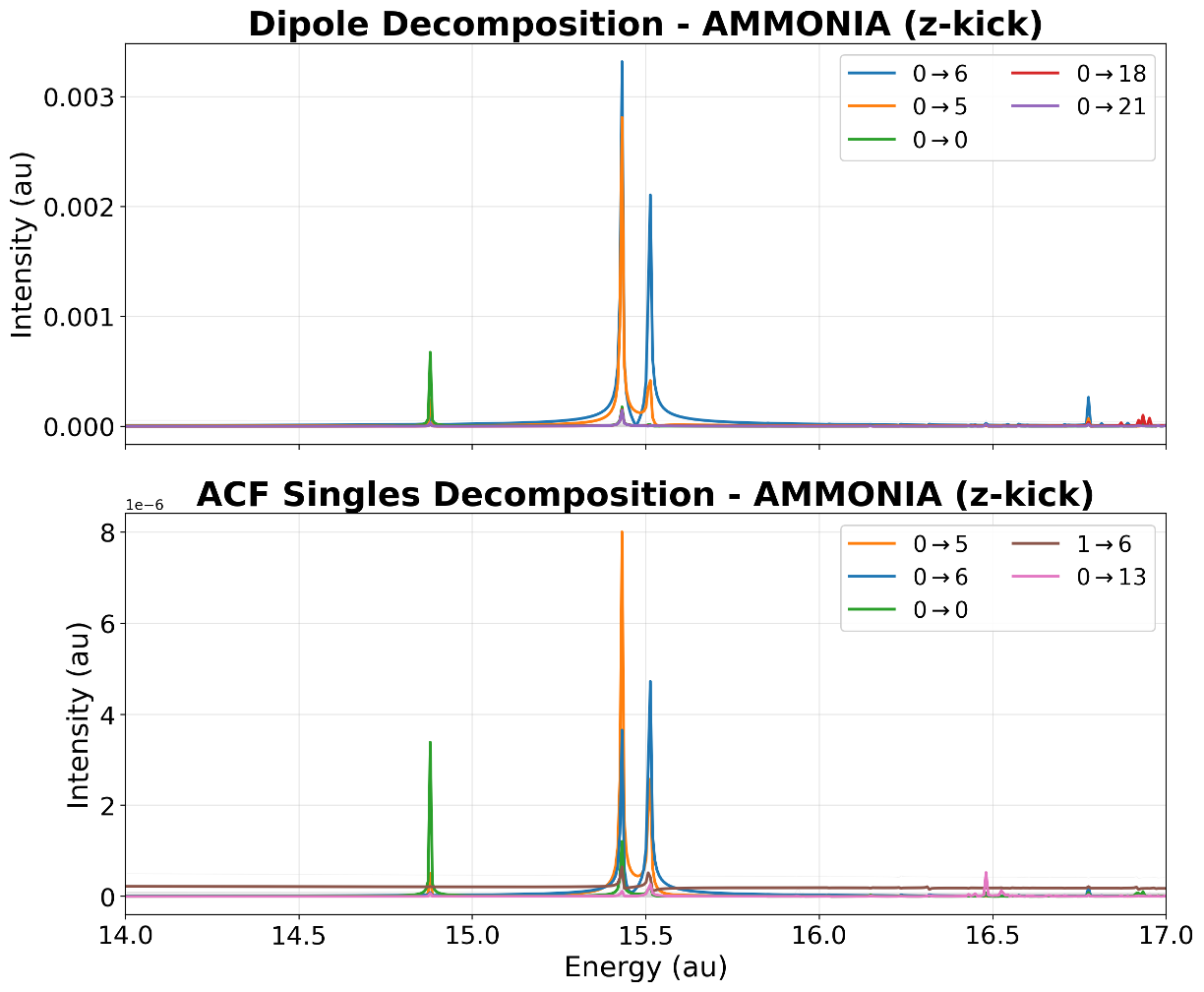}
    \caption{}
    \label{fig:amm_core}
  \end{subfigure}%
  \hspace*{\fill}%
  
  \captionsetup{skip=8pt}%
    \caption{Core spectral-region comparison of \ce{HF}, \ce{H2O} and \ce{NH3} under $z$-polarized perturbations at $0.01\,\mathrm{a.u.}$ field strength. Each panel shows dipole decomposition (top) and ACF singles decomposition (bottom) for the core-electron spectra. }
\label{fig:core}
\end{figure}
\subsection{Core-electron spectra}
A significant advantage of a time-dependent framework is its ability to treat all electronic excitations simultaneously through the use of a delta pulse, which generates a broadband spectrum. As a consequence, core-electron spectra emerge as a natural byproduct of the simulation rather than requiring specialized, frequency-targeted or other algorithms. Using our decomposition framework, we demonstrate this capability for \ce{HF}, \ce{H_2O} and \ce{NH3} in Fig.~\ref{fig:core} under $z$-polarized perturbations.

For these molecules, both dipole and ACF decompositions produce clean, well-resolved core spectra with consistent orbital assignments between the two methods, with the $1\mathrm{s}$ core transitions clearly identified and assigned to specific orbital-pair channels. The dominant transitions in each case reflect the same symmetry-determined selection rules observed in the valence region --- the $z$-polarized kick accesses only transitions of appropriate symmetry, depending on the molecule, orientation, and point group. These results show that the orbital-resolved interpretation of TD-CC dynamics extends seamlessly from the valence region into the X-ray regime, providing a unified description of molecular response. 

We note that a direct validation against EOM-CCSD, as performed for the valence transitions above, is not straightforward for core excitations within the standard EOM-CCSD framework. An accurate yet efficient treatment of core-excited states typically requires the core-valence separation (CVS) approximation to decouple valence and core excitations~\cite{Coriani2015}, and such a comparison is beyond the scope of the present work. While a more direct comparison might be possible using damped linear response theory and the asymmetric Lanczos algorithm \cite{Coriani2012}, the consistency between dipole and ACF decompositions nonetheless provides internal validation of the core-transition assignments. 

\subsection{Impulsive stimulated Raman X-ray scattering}

In a recent paper, \citeauthor{balbi_coupled_2023} used time-dependent EOM-CCSD (TD-EOM-CCSD) theory to simulate
impulsive stimulated Raman X-ray scattering (ISXRS) processes in atomic and molecular systems, including the ensuing
charge migration\cite{balbi_coupled_2023}. The ISXRS phenomenon can be explained qualitatively using electron configurations.
An attosecond or few-femtosecond laser pulse generates a core hole, either through bound core excitation or through core ionization,
while simultaneously stimulating a transition from a high-lying occupied valence orbital to the vacant core orbital.
Thus, the ISXRS process differs from Auger decay mainly by being stimulated by the laser field rather than spontaneous.
By expanding the time-dependent (right and left) state in a finite number of EOM-CCSD stationary states---which must be explicitly computed---the
interpretation of the laser-induced dynamics in terms of electron configurations can be straightforwardly done through analysis
of the EOM-CCSD eigenvectors.

If the qualitative picture of the ISXRS process remains valid in a correlated treatment, it should be clearly identifiable in terms of changes in the configuration weights
computed during a TD-CCSD simulation. To test this hypothesis, we choose the simplest possible model system, the \ce{Ne} atom, and expose it to
a laser pulse similar to the one used by \citeauthor{balbi_coupled_2023}\cite{balbi_coupled_2023}
Rather than using a Gaussian envelope which has
long tails that must be truncated, we use a trigonometric envelope with strictly finite support whilst remaining continuous at all times $t$,
\begin{equation}
    g_n(t) =
    \begin{cases}
        \cos^n\left( \frac{\pi(t - t_c)}{T_n} \right) & \vert t - t_c \vert \leq \frac{T_n}{2} \\
        0 & \vert t - t_c \vert > \frac{T_n}{2}
    \end{cases}
\end{equation}
Here, $n$ is an integer, $t_c$ is the central (peak) time of the envelope, and the foot-to-foot duration is given by
\begin{equation}
    T_n = \frac{\sqrt{\ln(2)}\pi \sigma}{\arccos\left(2^{-(2n)^{-1}}\right)}
\end{equation}
such that $g_n(t)$ rapidly converges to the Gaussian envelope 
$\exp(-(t-t_c)^2/(2\sigma^2))$ as $n \to \infty$\cite{Barth2009}.
The $z$-polarized time-dependent electric-field vector is then given by
\begin{equation}
    \label{eq:trig_laser}
    \boldsymbol{E}(t) = \boldsymbol{u}_z E_0 \cos\left(\omega (t-t_c) \right) g_n(t)
\end{equation}
where $\boldsymbol{u}_z$ is the unit vector along the $z$-axis, $E_0$ is the field strength, and $\omega$ the carrier frequency.
We use $t_c=0\,\text{a.u.}$, $n=10$, and choose the same standard deviation as \citeauthor{balbi_coupled_2023}, $\sigma = 5\,\text{a.u.}$\cite{note}
This gives $T_n = 49.961\,\text{a.u.} = 1.208\,\text{fs}$.
The field strength is $E_0 = 10\,\text{a.u.}$, corresponding to a peak intensity of $3\times 10^{18}\,\text{W/cm}^2$.
The carrier frequency $\omega=31.585717\,\text{E}_h=859.491145\,\text{eV}$
is taken from the supporting information of Ref.~\citenum{balbi_coupled_2023}.
\begin{figure}[h]
    \includegraphics{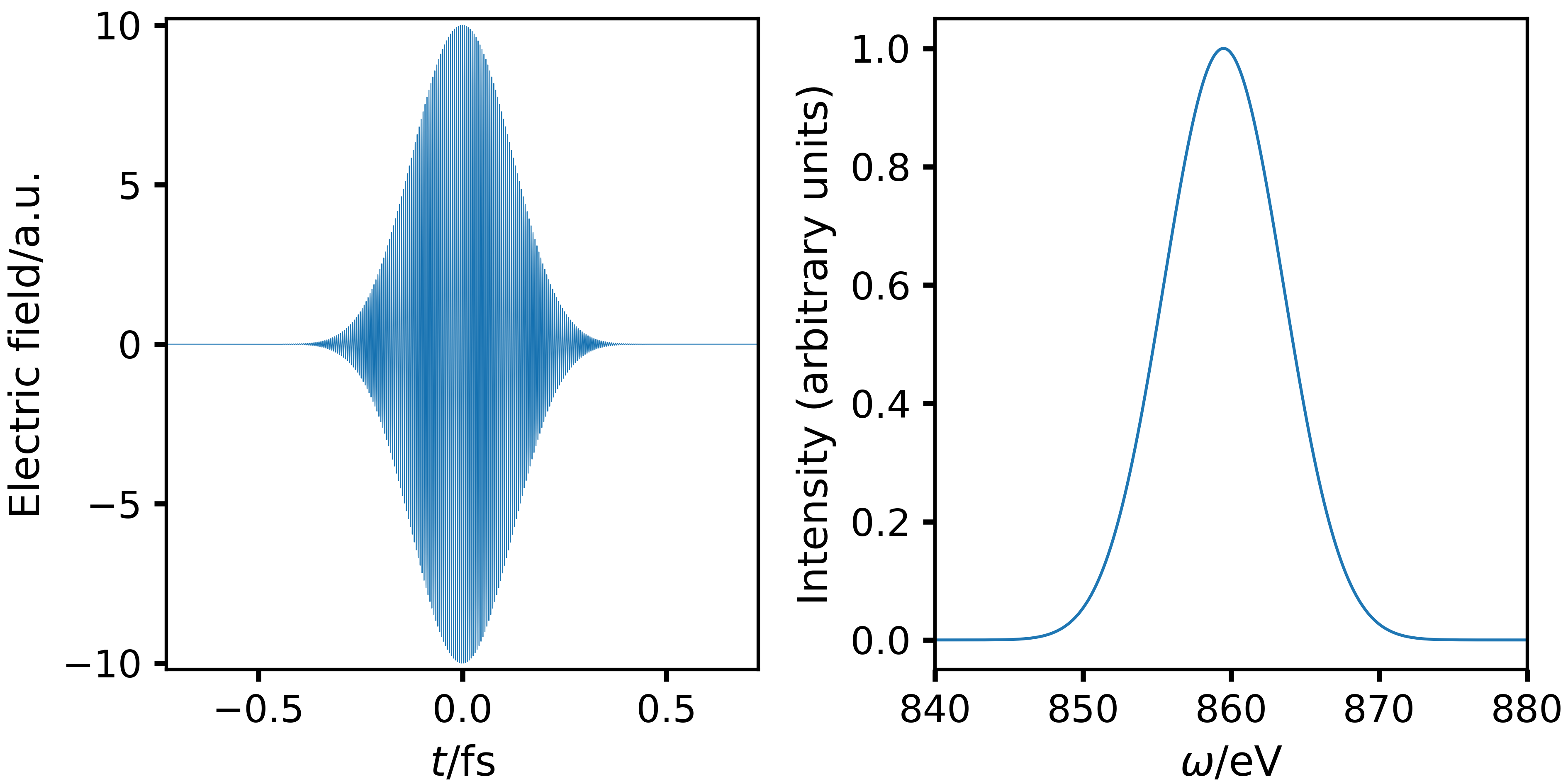}
    \caption{Laser pulse used to drive the ISXRS process in \ce{Ne}. Left: Electric-field amplitude as a function of time.
    Right: Intensity distribution computed as the absolute square of the Fourier transform of the electric-field amplitude.}
    \label{fig:isxrs_laser}
\end{figure}
For reference, the laser pulse is depicted in Fig.~\ref{fig:isxrs_laser} along with its intensity distribution. As can be verified
from the figure, the full-width-at-half-maximum of the intensity distribution is roughly $12\,\mathrm{eV}$ and thus broad enough
to induce several transitions during its brief interaction with the \ce{Ne} atom.

Using the CCSD ground state as initial condition, we run a TD-CCSD simulation with the aug-cc-pCVTZ\cite{Dunning1989,Kendall1992,Woon1995}
basis set, recording configuration weights at each time step. 
The TD-CCSD state is propagated from $t = -30\,\text{a.u.} = -0.726\,\text{fs}$ to
$t = 30\,\text{a.u.} = 0.726\,\text{fs}$ in time steps of $\Delta t = 0.005\,\text{a.u.}$.
We use the sixth-order Gauss-Legendre integrator as described in Ref.~\citenum{pedersen2019symplectic}, with tolerance
$10^{-7}$ in the fixed-point iterations, as implemented in the HyQD software library~\cite{HyQD}. Hamiltonian integrals
and the HF ground state are computed using the PySCF package~\cite{pyscf,pyscf2}. The HF and CCSD ground-state optimizations are
tightly converged using a tolerance of $10^{-10}$ for the HF orbital gradient and the CCSD residual norms, respectively.

\begin{figure}[h]
    \includegraphics{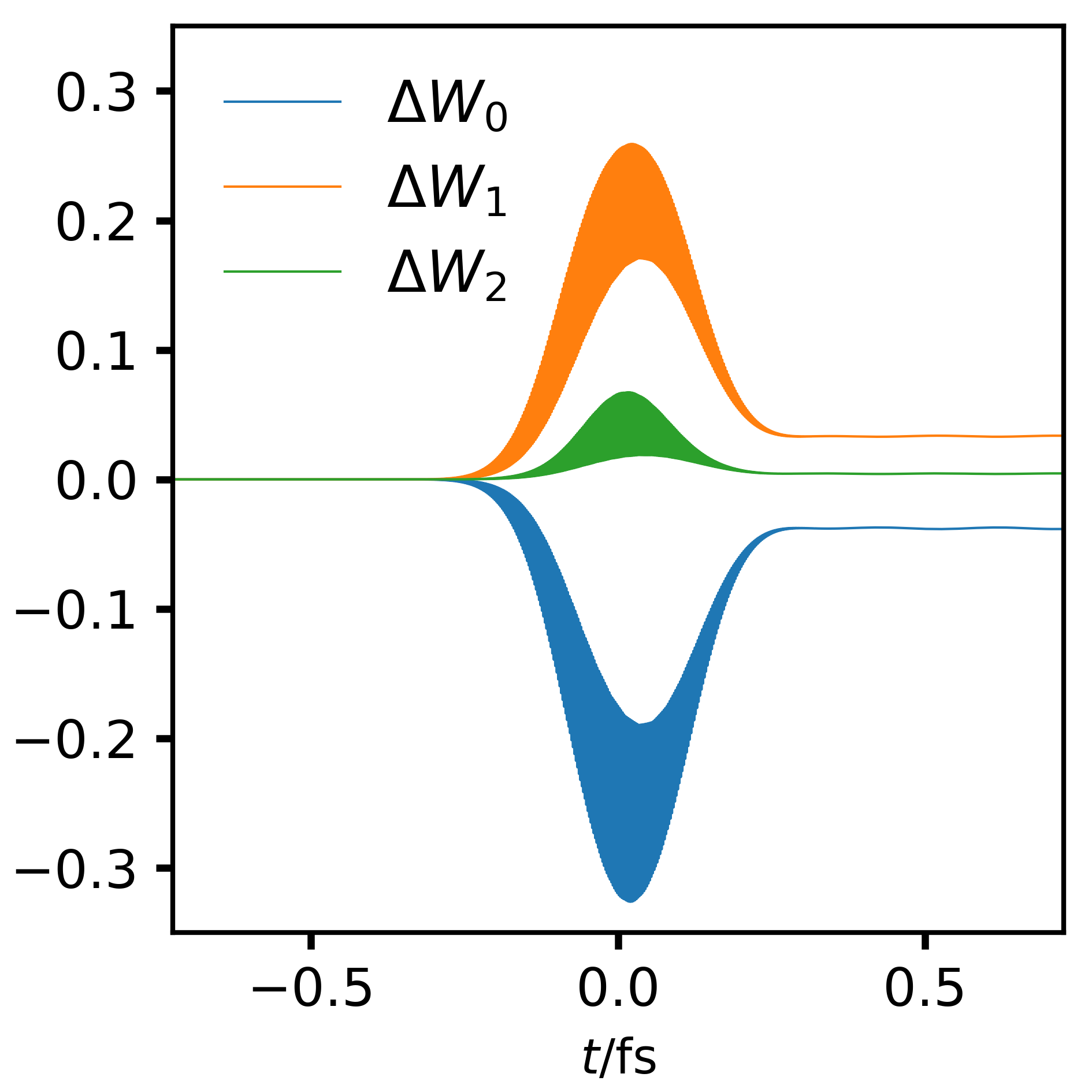}
    \caption{Time evolution of the CCSD reference, singles, and doubles configuration weights during an ISXRS process in the \ce{Ne}
    atom, obtained with the aug-cc-pCVTZ basis set. The weights are plotted relative to their ground-state values.}
    \label{fig:isxrs_ne_aug-cc-pcvtz_total_weights}
\end{figure}

The reference determinant is the HF ground state $\ket{\Phi_0} = \ket{1\text{s}^22\text{s}^22\text{p}^6}$, which has weight
$W_0 = 0.96366$ in the CCSD ground state. The total singles and doubles weights of the ground state are 
$W_1 = 0.00057$ and $W_2 = 0.03577$, respectively. The changes in these weights during the dynamics are shown in 
Fig.~\ref{fig:isxrs_ne_aug-cc-pcvtz_total_weights}.
The reference weight drops by more than $30\%$ with a concomitant increase in the total singles weight while the correlating doubles weight
remains relatively low. At the end of the simulation, the weights almost return to their ground-state values. Only the weights of a few single-excited
determinants deviate noticeably from their ground-state values, despite the very high intensity of the laser pulse. It should be recalled, however,
that the chosen Gaussian basis functions do not provide a reasonable representation of the electronic continuum and, hence, ionization
processes are not properly captured in the simulation. We note in passing that replacing atom-centered fixed Gaussians with 
fully flexible Gaussian wave packets would correct this flaw.~\cite{kvaal_no_2023,schrader_time_2024,wozniak_gaussians_2024,schrader_multidimensional_2025,wozniak_rothe_2025,schrader_time-dependent_2025}

\begin{figure}[h]
    \includegraphics{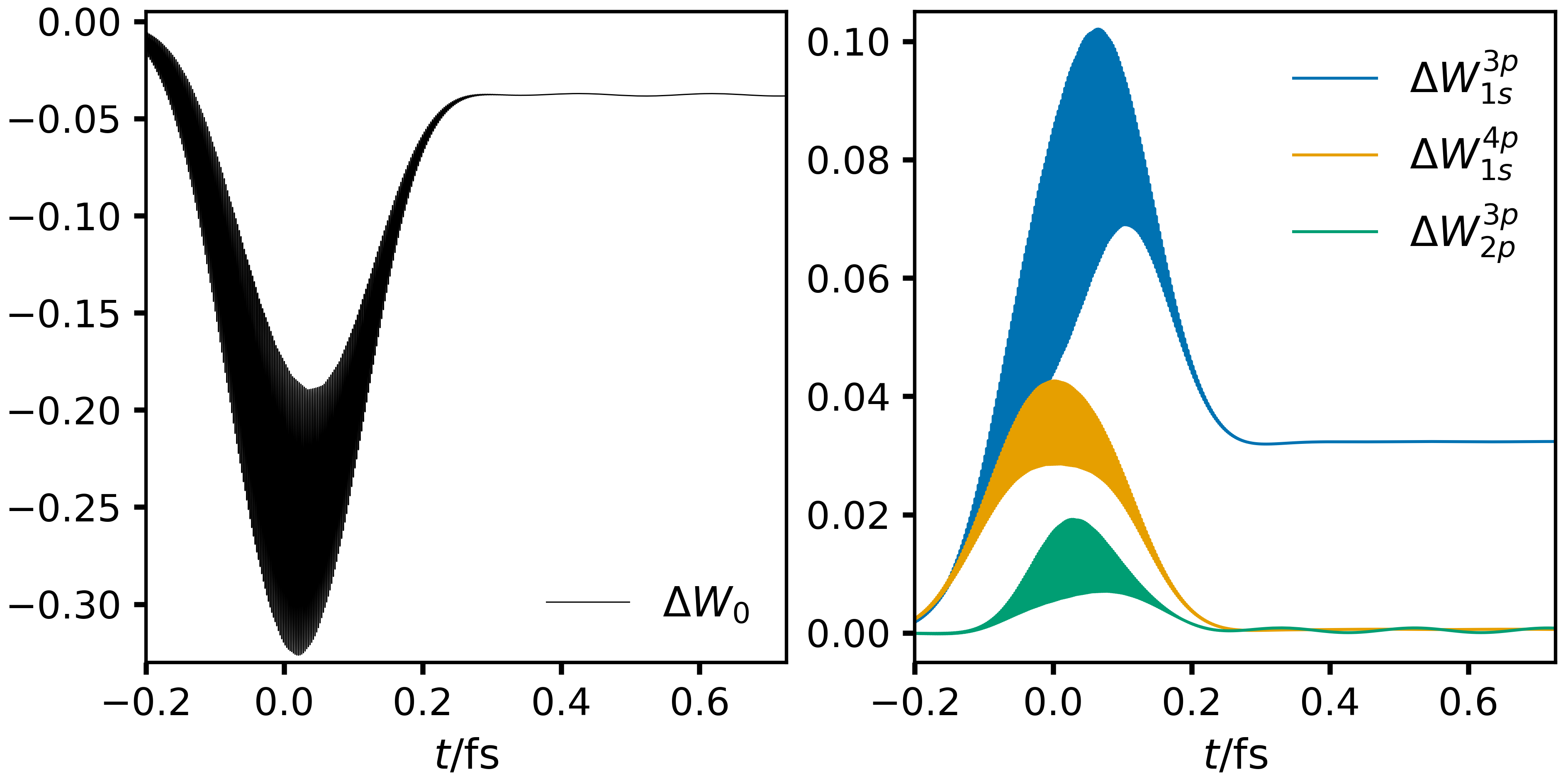}
    \caption{Time evolution of key CCSD configuration weights involved in an ISXRS process in the \ce{Ne}
    atom, obtained with the aug-cc-pCVTZ basis set. The weights are plotted relative to their ground-state values.}
    \label{fig:isxrs_ne_aug-cc-pcvtz}
\end{figure}

Figure \ref{fig:isxrs_ne_aug-cc-pcvtz} shows changes relative to the ground-state values of the reference weight and all singles weights
contributing at least $0.5\%$ of the maximum singles weight after the laser pulse is switched off.
A considerably greater number of single-excited determinants contribute to the dynamics during the matter-field interaction (cf.~Fig.~\ref{fig:isxrs_ne_aug-cc-pcvtz_total_weights}),
but only the ones selected show significant changes relative to their ground-state values after the interaction.
Initially, the laser pulse induces transitions from the $1\mathrm{s}$
core orbital to the unoccupied $3\mathrm{p}$ and $4\mathrm{p}$ orbitals, populating the $\ket{1\text{s}^12\text{s}^22\text{p}^63\text{p}^1}$
and $\ket{1\text{s}^12\text{s}^22\text{p}^64\text{p}^1}$ single-excited determinants. Once these are sufficiently populated, at $t\approx -0.1\,\text{fs}$,
the $\ket{1\text{s}^22\text{s}^22\text{p}^53\text{p}^1}$ determinant becomes populated through de-excitation from the $2\mathrm{p}$ valence orbital
to the $1\mathrm{s}$ core orbital, i.e., through the transition from $\ket{1\text{s}^12\text{s}^22\text{p}^63\text{p}^1}$ to
$\ket{1\text{s}^22\text{s}^22\text{p}^53\text{p}^1}$.
Note that $\ket{1\text{s}^22\text{s}^22\text{p}^53\text{p}^1}$ represents a dark state,
i.e., it  is electric-dipole forbidden and hence cannot be populated through direct excitation from the HF ground-state determinant.

\begin{figure}[h]
    \includegraphics{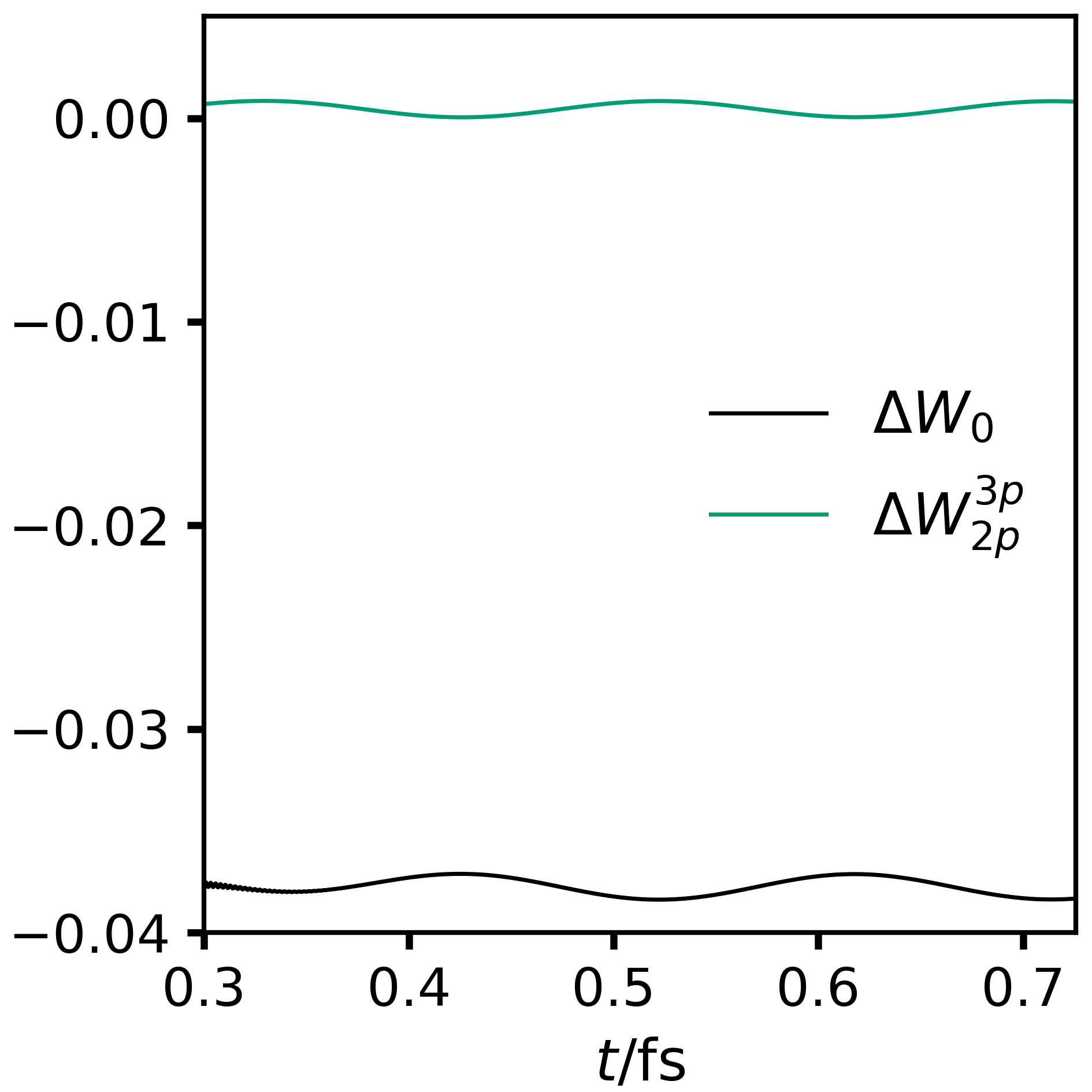}
    \caption{Field-free coupling of the HF ground-state determinant and the dipole-forbidden $\ket{1\text{s}^22\text{s}^22\text{p}^53\text{p}^1}$
    determinant of \ce{Ne}, following an ISXRS process simulated with the TD-CCSD method and the aug-cc-pCVTZ basis set. 
    }
    \label{fig:isxrs_ne_aug-cc-pcvtz_zoom}
\end{figure}

The final population of $\ket{1\text{s}^22\text{s}^22\text{p}^53\text{p}^1}$ is small, though.
Moreover, the $\ket{1\text{s}^22\text{s}^22\text{p}^53\text{p}^1}$ population weakly oscillates after the pulse has been switched off,
whereas the $\ket{1\text{s}^12\text{s}^22\text{p}^63\text{p}^1}$ and $\ket{1\text{s}^12\text{s}^22\text{p}^64\text{p}^1}$ populations are
very nearly constant. The weak oscillations are caused by field-free coupling to the ground state determinant whose population
shows almost synchronous oscillations, as seen in Fig.~\ref{fig:isxrs_ne_aug-cc-pcvtz_zoom}.

\subsection{Transient absorption}

Ultrafast transient absorption spectroscopy allows us to study electron dynamics on its natural time scale using
a pump pulse to excite electrons from the ground state and subsequently, at varying time delays, probing the generated
electronic wave packet using a second pulse, the probe pulse\cite{Wu2016}. The first TD-CCSD simulations of transient
absorption spectra were published in \citeyear{skeidsvoll2020time} by \citeauthor{skeidsvoll2020time}\cite{skeidsvoll2020time},
who studied optical pump -- X-ray probe spectra of \ce{LiH} and \ce{LiF}. The time-resolved spectra were assigned by means
of EOM-CCSD eigenstates identified by their energies in the time-resolved pump-probe spectra, not from the simulations themselves.
In this work, we study a pump-probe spectrum of the \ce{HF} molecule, using only simulation data for interpretation.
The fixed \ce{HF} bond distance is $0.928663\,\text{\AA}$ and the molecule is placed on the $z$-axis.

Using the aug-cc-pCVDZ basis set (i.e., aug-cc-pCVDZ for \ce{F} and aug-cc-pVDZ for \ce{H}),~\cite{Dunning1989,Kendall1992,Woon1995}
the linear absorption spectrum obtained from a TD-CCSD simulation with an electric-field kick polarized along the
molecular ($z$-) axis exhibits a valence absorption at approximately $14.36\,\text{eV}$ and a core excitation at roughly $695.6\,\text{eV}$.
The ground-state configuration is $\ket{\Phi_0} = \ket{1\sigma^22\sigma^23\sigma^21\pi^4}$ and
a dipole decomposition shows that about half of the intensity of the valence transition stems from the $3\sigma \to 4\sigma$ excitation, with
the remaining intensity stemming roughly equally from the $3\sigma \to 5\sigma$ and $3\sigma \to 6\sigma$ excitations.
The intensity of the core transition is about $50\%$ $1\sigma \to 5\sigma$, with the other half distributed
roughly equally among the $1\sigma \to 4\sigma$, $1\sigma \to 6\sigma$, $1\sigma \to 7\sigma$, and $1\sigma \to 8\sigma$ excitations.
Note, however, that the intensity contribution from $1\sigma \to 4\sigma$ is negative (i.e., one might perceive it
as the de-excitation, $4\sigma \to 1\sigma$).

A simplified configuration-based picture of these transitions thus involves the space spanned by the valence-excited determinants
$\mathcal{V}_\mathrm{v} = \{ \ket{\Phi_{3\sigma}^{4\sigma}}, \ket{\Phi_{3\sigma}^{5\sigma}}, \ket{\Phi_{3\sigma}^{6\sigma}}\}$
and the space spanned by the core-excited determinants
$\mathcal{V}_\mathrm{c} = \{ \ket{\Phi_{1\sigma}^{4\sigma}}, \ket{\Phi_{1\sigma}^{5\sigma}}, \ket{\Phi_{1\sigma}^{6\sigma}}, \ket{\Phi_{1\sigma}^{7\sigma}}, \ket{\Phi_{1\sigma}^{8\sigma}}\}$, in addition to the reference determinant $\ket{\Phi_0}$.
The determinants within each of the subspaces $\mathcal{V}_\mathrm{v}$ and $\mathcal{V}_\mathrm{c}$ are strongly dipole coupled, with off-diagonal dipole 
matrix elements ranging from $0.2$ to $2.2\,\mathrm{a.u.}$ within $\mathcal{V}_\mathrm{v}$ and from $0.9$ to $2.2\,\mathrm{a.u.}$ within $\mathcal{V}_\mathrm{c}$. The dipole coupling between $\mathcal{V}_\mathrm{v}$ and $\mathcal{V}_\mathrm{c}$ is non-zero but about two orders of magnitude smaller, implying
that weak weight (population) transfer between the two subspaces may be induced in a pump-probe setup. In addition, the determinants will
continue interacting after the matter-field interaction due to the fluctuation potential, i.e., due to electron correlation.

To investigate these interactions, we run TD-CCSD simulations with the CCSD ground state as initial state,
exposing the \ce{HF} molecule to two consecutive laser pulses of the form \eqref{eq:trig_laser} with $n=10$. 
The parameters of the pump laser are:
$E_0 = 0.01\,\text{a.u.}$,
$\omega = 0.5277\,\text{a.u.} = 14.36\,\text{eV}$,
$T_n = 199.844\,\text{a.u.} = 4.834\,\text{fs}$,
and
$t_c = 0\,\text{a.u.}$
The parameters of the probe laser are:
$E_0 = 0.1\,\text{a.u.}$,
$\omega = 25.5628\,\text{a.u.} = 695.6\,\text{eV}$,
$T_n = 59.953\,\text{a.u.} = 1.450\,\text{fs}$,
and
$t_c = \Delta t_c$.
The delay between the pump and probe lasers, $\Delta t_c$, is varied in the interval from $-130\,\text{a.u.}$ to
$225\,\text{a.u.}$ in steps of $5\,\text{a.u.}$.
The TD-CCSD state is propagated from $t = -1000\,\text{a.u.} = -24.19\,\text{fs}$ to
$t = 1000\,\text{a.u.} = 24.19\,\text{fs}$ in time steps of $\Delta t = 0.01\,\text{a.u.}$. Note that the time steps before the first laser is switched on
can be performed without actual calculation, as the field-free ground-state propagation only gives rise to a trivial
phase factor~\cite{pedersen2019symplectic}.
We use the fourth-order Gauss-Legendre integrator with tolerance
$10^{-7}$ in the fixed-point iterations, as implemented in the HyQD software library~\cite{HyQD}. Hamiltonian integrals
and the molecular orbitals are computed using the PySCF package~\cite{pyscf,pyscf2}. The CCSD ground-state optimization is
tightly converged using a tolerance of $10^{-10}$, including the orbital gradient in the preceding molecular-orbital calculation.
For reference, we run two additional simulations, one with only the pump laser and another with only the probe laser, using $t_c = 0\,\text{a.u.}$
in both cases.

\begin{figure}[htb]
    \includegraphics[scale=0.5]{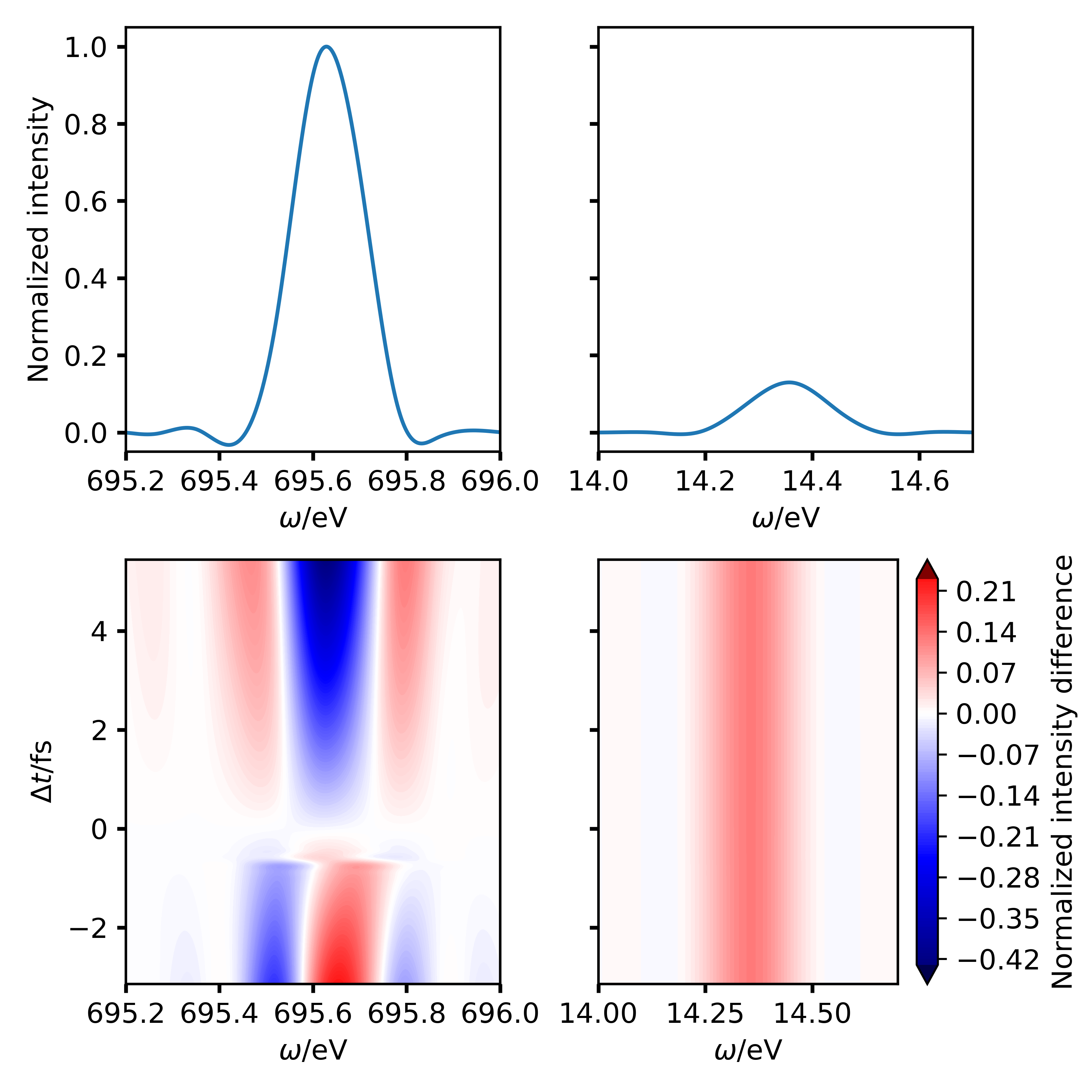}
    \caption{Optical pump -- X-ray probe transient absorption spectrum of \ce{HF}, computed from
    TD-CCSD theory with the aug-cc-pCVDZ basis set.
    Top left: normalized core-region probe spectrum.
    Top right: valence-region pump spectrum normalized with respect to the probe spectrum.
    Bottom left: core-region difference spectrum at a range of pump--probe delays normalized with respect to the probe spectrum.
    Bottom right: valence-region difference spectrum at a range of pump--probe delays normalized with respect to the probe spectrum.
    Cubic spline interpolation has been applied in all cases to create a smooth plot.}
    \label{fig:HF_aug-cc-p_c_vdz_pump_probe}
\end{figure}

The resulting pump-probe spectrum is shown in Fig.~\ref{fig:HF_aug-cc-p_c_vdz_pump_probe} along with the reference pump and probe spectra.
The probe spectrum is plotted after normalization, i.e.,
\begin{equation}
    \omega S_\mathrm{probe}(\omega) \leftarrow \mathcal{N}_\mathrm{probe} \omega S_\mathrm{probe}(\omega)
\end{equation}
where the normalization constant is defined as
\begin{equation}
    \mathcal{N}_\mathrm{probe} = \max_\omega \omega S_\mathrm{probe}(\omega)
\end{equation}
The pump spectrum is normalized with respect to the probe spectrum,
\begin{equation}
    \omega S_\mathrm{pump}(\omega) \leftarrow \mathcal{N}_\mathrm{probe} \omega S_\mathrm{pump}(\omega)
\end{equation}
The pump-probe spectrum, both in the core region and in the valence region, is plotted relative to the probe spectrum,
\begin{equation}
    \omega S_\mathrm{pump-probe}(\omega) \leftarrow \mathcal{N}_\mathrm{probe} \omega \left( S_\mathrm{pump-probe}(\omega) - S_\mathrm{probe}(\omega)\right)
\end{equation}
where the probe spectrum on the right-hand side is not normalized.
The fast Fourier transform is used to obtain the spectral function from the time-dependent electric-dipole moment after multiplication with the
Hann window function, $\cos^2( \pi t / 2000 )$.

While the valence absorption is only slightly enhanced by the probe laser with little sensitivity to the delay, the core absorption shows delay-dependent
oscillations with significant amplitude. Depending on the delay, the central peak at $695.6\,\mathrm{eV}$ is decreased by up to $40\%$ at large delays
and increased by up to $20\%$ at negative delays (i.e., when the probe laser is applied \emph{before} the pump laser).
To gain insights into the electron dynamics underpinning these features, we compute the total weights of the determinants in
$\mathcal{V}_\mathrm{v}$ and $\mathcal{V}_\mathrm{c}$.

\begin{figure}[htb]
    \includegraphics[scale=0.75]{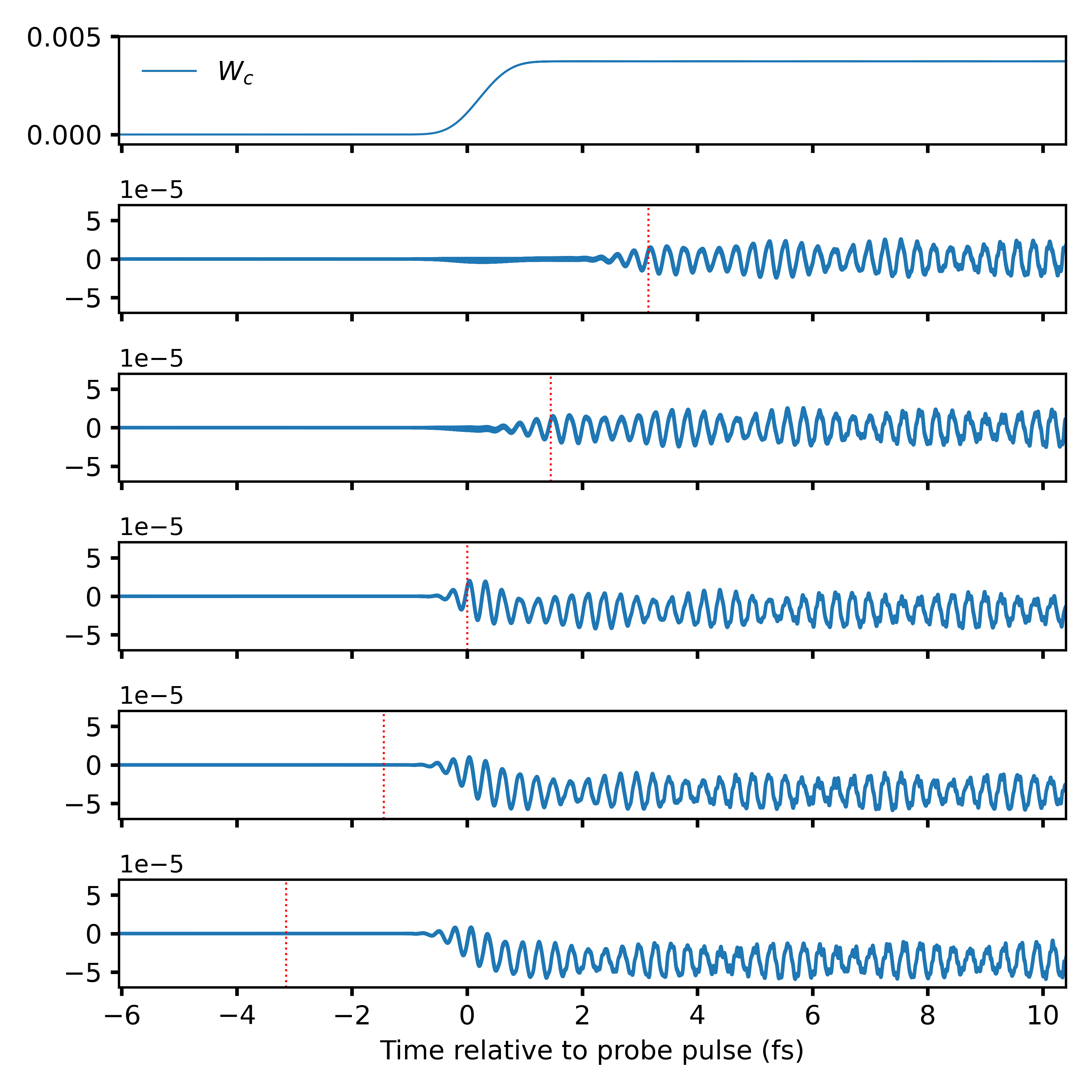}
    \caption{Total TD-CCSD weight of the single-excited determinants in $\mathcal{V}_\mathrm{c}$ as a function of time relative to the central time of the
    probe pulse. The top panel shows the weight when only the probe pulse is applied, while the panels below show the changes in the same trajectory when
    both the pump and probe pulses are applied, with varying delay. The vertical dotted lines indicate the central time of the pump pulse
    (relative to the probe pulse).
    The pump-probe delays shown are $-130$, $-60$, $0$, $60$, and $130\,\text{a.u.}$, counting from top to bottom.}
    \label{fig:hf_w_c}
\end{figure}

Figure \ref{fig:hf_w_c} shows the time evolution of the total weight of determinantes in $\mathcal{V}_\mathrm{c}$ during and after the interaction with
the probe pulse, with and without the pump pulse at different delays.
The pump pulse introduces small changes of about $1\%$ with oscillations that persist after it is switched off,
indicating that electron correlation is the main driving force of the interaction between valence-excited and core-excited determinants,
causing the intensity changes in Fig.~\ref{fig:HF_aug-cc-p_c_vdz_pump_probe}. Moreover, changes caused by the pump pulse
clearly depend on the delay, with a reduction in weight when the probe pulse is applied after the pulse in qualitative agreement with
the reduced intensity of the central peak at positive delays in Fig.~\ref{fig:HF_aug-cc-p_c_vdz_pump_probe}.

\begin{figure}[htb]
    \includegraphics[scale=0.75]{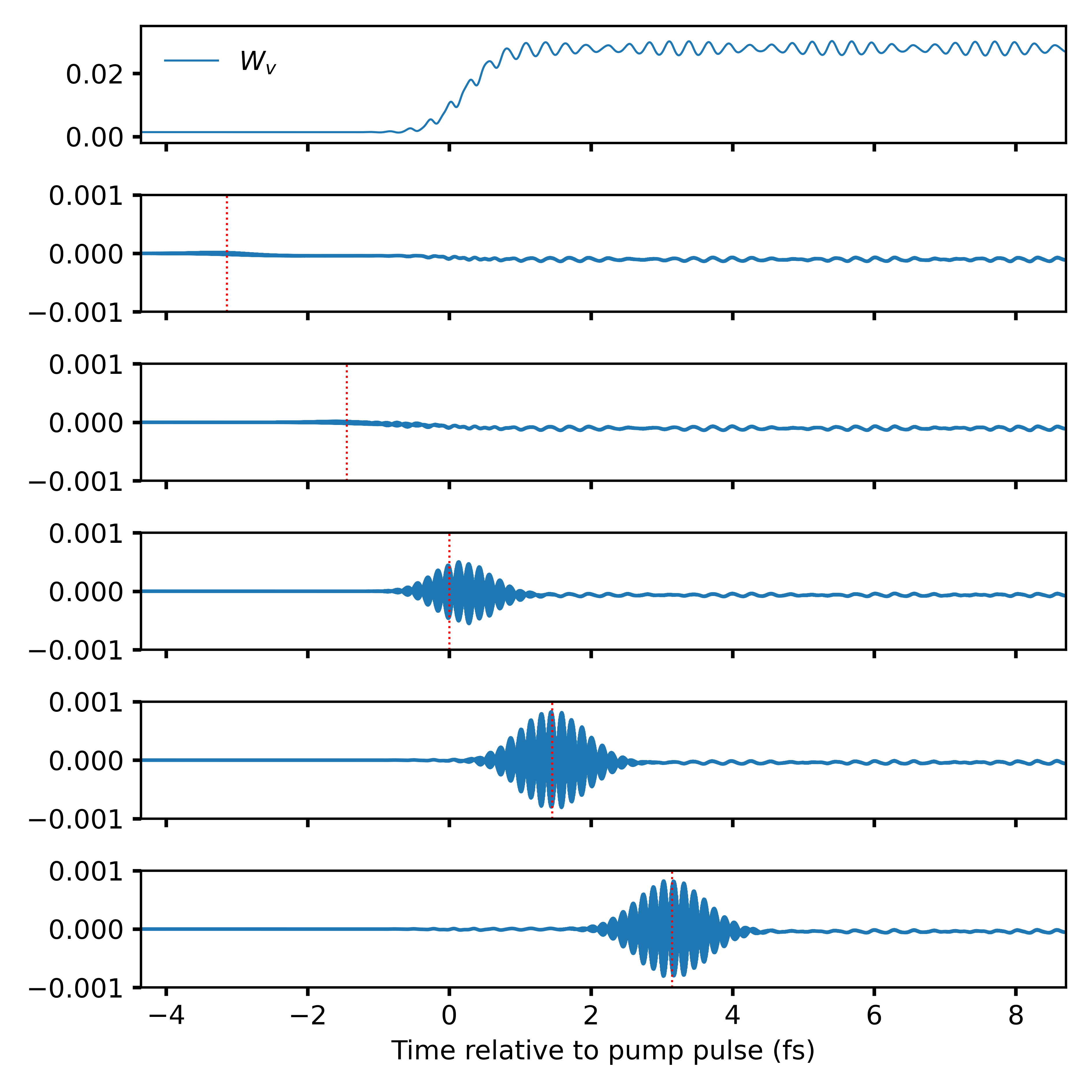}
    \caption{Total TD-CCSD weight of the single-excited determinants in $\mathcal{V}_\mathrm{v}$ as a function of time relative to the central time of the
    pump pulse. The top panel shows the weight when only the pump pulse is applied, while the panels below show the changes in the same trajectory when
    both the pump and probe pulses are applied, with varying delay. The vertical dotted lines indicate the central time of the probe pulse
    (relative to the pump pulse).
    The pump-probe delays shown are $-130$, $-60$, $0$, $60$, and $130\,\text{a.u.}$, counting from top to bottom.}
    \label{fig:hf_w_v}
\end{figure}

Conversely, the time evolution of the weight of the valence excited determinants in $\mathcal{V}_\mathrm{v}$ displayed in
Fig.~\ref{fig:hf_w_v} shows that the probe pulse has limited effect.
While the probe pulse introduces weight changes of up to about $4\%$ during the matter-field interaction, the weight largely returns to its
initial value after the interaction. This is in qualitative agreement with the delay-independent intensity of the central valence peak in Fig.~\ref{fig:HF_aug-cc-p_c_vdz_pump_probe}.

\begin{figure}[htb]
    \includegraphics[scale=0.75]{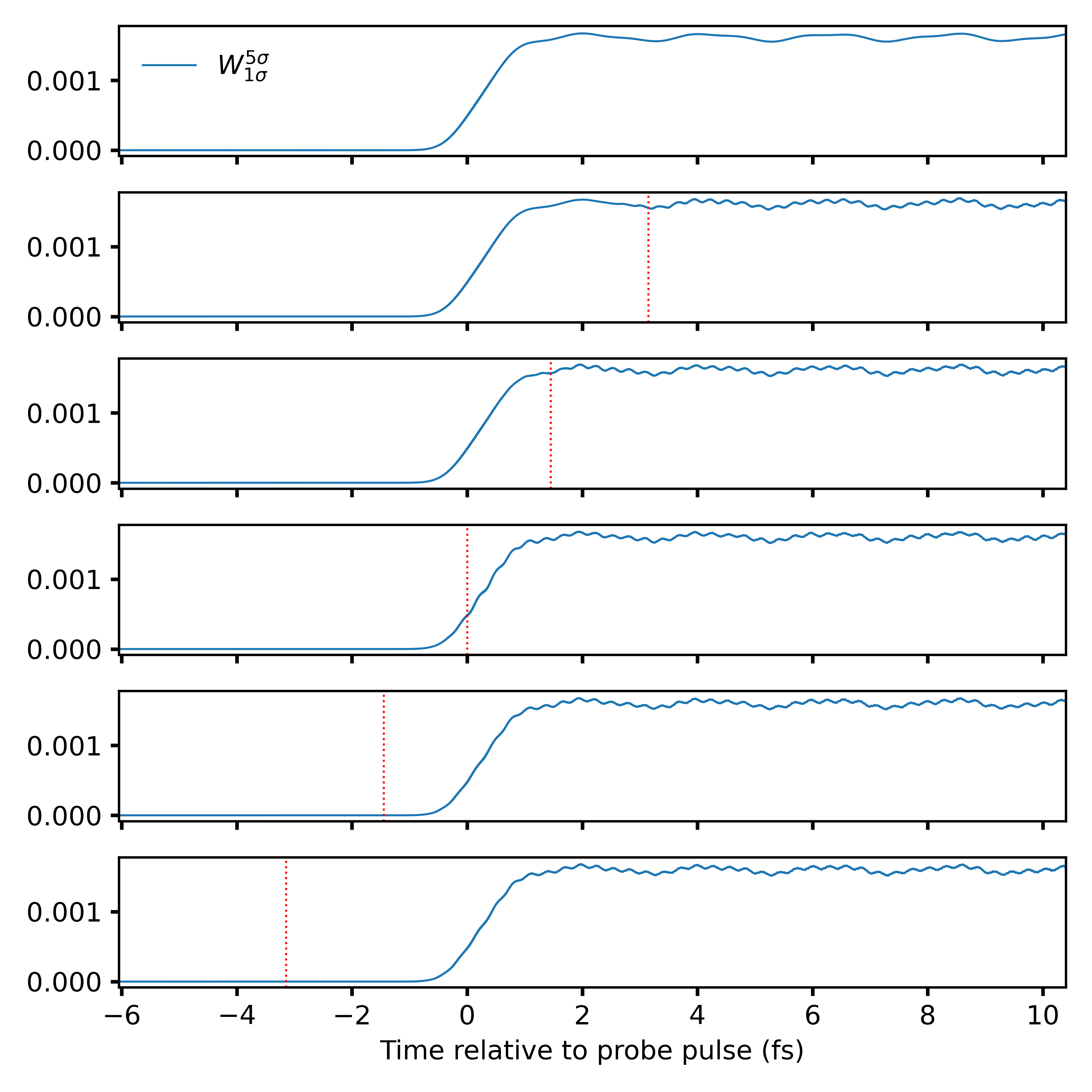}
    \caption{TD-CCSD weight of the core-excited determinant $\ket{\Phi_{1\sigma}^{5\sigma}}$ as a function of time relative to the central time of the
    probe pulse. The top panel shows the weight when only the probe pulse is applied, while the panels below show the weight when
    both the pump and probe pulses are applied, with varying delay. The vertical dotted lines indicate the central time of the pump pulse
    (relative to the probe pulse).
    The pump-probe delays shown are $-130$, $-60$, $0$, $60$, and $130\,\text{a.u.}$, counting from top to bottom.}
    \label{fig:hf_pump-probe_0_1}
\end{figure}

\begin{figure}[htb]
    \includegraphics[scale=0.75]{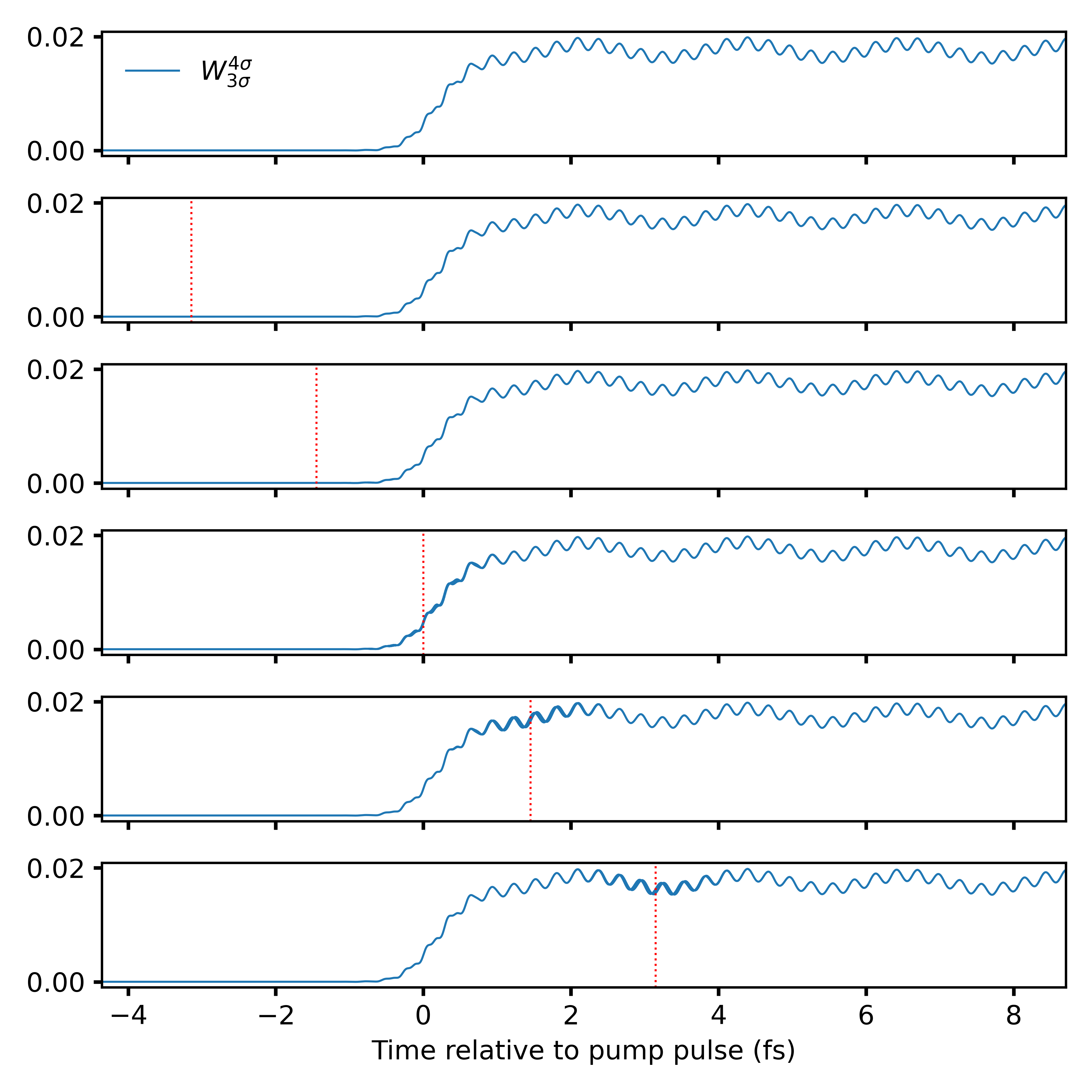}
    \caption{TD-CCSD weight of the valence-excited determinant $\ket{\Phi_{3\sigma}^{4\sigma}}$ as a function of time relative to the central time of the
    pump pulse. The top panel shows the weight when only the pump pulse is applied, while the panels below show the weight when
    both the pump and probe pulses are applied, with varying delay. The vertical dotted lines indicate the central time of the probe pulse
    (relative to the pump pulse).
    The pump-probe delays shown are $-130$, $-60$, $0$, $60$, and $130\,\text{a.u.}$, counting from top to bottom.}
    \label{fig:hf_pump-probe_2_0}
\end{figure}

These features can also be observed in the weights of individual single-excited determinants, as shown for the dominant
core- and valence-excited determinants $\ket{\Phi_{1\sigma}^{5\sigma}}$ and $\ket{\Phi_{3\sigma}^{4\sigma}}$
in Fig.~\ref{fig:hf_pump-probe_0_1} and \ref{fig:hf_pump-probe_2_0}, respectively.

\section{Conclusion} \label{sec:conc}

Based on expansions of the CC bra and ket functions in Slater determinant basis, we have presented a theoretical framework for extracting
chemical insight at the orbital level directly from real-time TD-CC simulations with negligible computational overhead.
We have illustrated the framework for time-dependent configuration weights and orbital decompositions of the time-dependent electric dipole moment
and of the autocorrelation function. While configuration weights provide direct insights into the electron dynamics in the time domain, the dipole and
ACF decompositions are useful for assigning orbital transitions in the energy domain.

For linear absorption spectra,
we have presented a systematic comparison of dipole and ACF decomposition methods for orbital-resolved spectroscopy, validated against time-independent (conventional) EOM-CCSD
results for four ten-electron molecules belonging to the $D_{\infty h}$, $C_{2v}$, $C_{3v}$, and $T_d$ point groups.
Both methods successfully identify electronic single-excitation characters, with normalized ACF amplitudes showing good agreement with EOM-CCSD eigenvector components (typical deviations $0.01$-$0.02$) despite measuring fundamentally different quantities: time-evolved amplitudes versus static eigenvector components.
This establishes the reliability of both decomposition approaches for orbital-resolved TD-CC simulations of linear absorption spectroscopy. The broadband character of the delta-pulse perturbation further allowed us to demonstrate that both decomposition methods extend seamlessly from the valence region into the X-ray regime, with the core-electron transitions in these molecules emerging as natural byproducts of the same simulations used for valence spectral analysis. 

For high-symmetry systems or non-degenerate states, dipole decomposition provides reliable spectral assignment, while ACF reveals complementary population dynamics.
For ultrafast phenomena such as impulsive stimulated X-ray Raman scattering and transient pump-probe spectroscopy, the dipole and ACF decompositions can be
complemented by time-dependent configuration weights to unveil the underlying processes at the orbital level, as illustrated here by computational investigations of ISXRS in the \ce{Ne} atom and optical pump -- X-ray probe spectroscopy of the \ce{HF} molecule. The qualitative configuration-based account of the ISXRS process
is clearly identifiable through the time-dependent weights, although multiple other configurations play an active role during the matter-field interaction.
Similarly, intensity variations in pump-probe spectra as a function of delay can be interpreted by studying the effect of the laser pulses on
configuration weights as functions of time, both before, during, and after the interaction.

Finally, we stress that although we have only considered TD-CCSD theory in the present work, the decompositions and configuration weights
can be easily extended to other truncations of the cluster operators, and to TD-CC models that include higher-order excitations in an approximate manner through perturbation theory~\cite{configurationwts}.

\section*{Supporting Information}
Cartesian coordinates (\ce{HF}, \ce{H2O}, \ce{NH3}, \ce{CH4}).
\section*{Acknowledgement}
This work was supported by the Research Council of Norway through its Centres of Excellence scheme, Project no.~262695 and
by Sigma2---the National Infrastructure for High Performance Computing and Data Storage in Norway, Grant No.~NN4654K (HEK, TBP).  AK and TDC were supported by the U.S.\ National Science Foundation via grant CHE-2154753.

\bibliography{refs}

\end{document}